\documentclass[aps,prb,twocolumn,showpacs,10pt,floatfix,citeautoscript,longbibliography]{revtex4-1}
\usepackage{amssymb}
\usepackage{amsmath}
\usepackage[pdftex]{graphicx}
\usepackage{dcolumn}
\usepackage{bm}
\usepackage{textcomp}

\usepackage{xcolor}
\definecolor{trueblue}{rgb}{0.0, 0.45, 0.81}
\definecolor{crimsonglory}{rgb}{0.75, 0.0, 0.2}

\usepackage{hyperref}
\hypersetup{colorlinks,linkcolor={trueblue},citecolor={crimsonglory},urlcolor={crimsonglory}}

\begin{document}

\title{Low-energy theory for strained graphene: an approach up to second-order in the strain tensor}


\author{M. Oliva-Leyva}
\email{moliva@iim.unam.mx}
\author{Chumin Wang}
\email{chumin@unam.mx}

\affiliation{Instituto de Investigaciones en Materiales, Universidad Nacional Aut\'{o}noma de
M\'{e}xico, Apartado Postal 70-360, 04510 Mexico City, Mexico.}


\begin{abstract}

An analytical study of low-energy electronic excited states in an uniformly strained graphene is carried out up to second-order in the strain tensor. We report an new effective Dirac Hamiltonian with an anisotropic Fermi velocity tensor, which reveals the graphene trigonal symmetry being absent in low-energy theories to first-order in the strain tensor. In particular, we demonstrate the dependence of the Dirac-cone elliptical deformation on the stretching direction respect to graphene lattice orientation. We further analytically calculate the optical conductivity tensor of strained graphene and its transmittance for a linearly polarized light with normal incidence. Finally, the obtained analytical expression of the Dirac point shift allows a better determination and understanding of pseudomagnetic fields induced by nonuniform strains.

\end{abstract}


\maketitle

\section{Introduction}

Given the striking interval of elastic response of graphene \cite{Galiotis2015,Colin2015}, can withstand a reversible stretching up to $25\,\%$, strain engineering has been widely used to improve and/or to tune its electronic, thermal, chemical and optical properties \cite{Bissett2014,Zhang2015,Si2016,Amorim2016,Oliva2017}. For instance, theoretical predictions have been made of a band-gap opening by large uniaxial strains from both tight-binging approach \cite{Pereira09a} and density functional theory \cite{Choi2010}, whenever the strain produces such a Hamiltonian modification beyond the inequalities obtained by Hasegawa, \emph{et al}. \cite{Hasegawa} The emergence of the pseudomagnetic field caused by a nonuniform strain is possibly the most interesting strain-induced electronic effect, due to the possibility of observing a pseudoquantum Hall effect under zero external magnetic fields \cite{Guinea2010a,Guinea2010b}. Nowadays, the transport signatures of the such fictitious fields are actively investigated \cite{Falko2013,Bahamon2015,Burgos2015,Mikkel16a,Stegmann2016,Mikkel16b,Carrillo2016,Georgi2016}. Moreover, from a view point of basic research, strained graphene opens an opportunity to explore mixed Dirac--Schr\"odinger Hamtiltonian \cite{deGail2012}, fractal spectrum \cite{Naumis2014}, superconducting states \cite{Kauppila2016}, magnetic phase transitions \cite{Brey2016}, metal-insulator transition \cite{Tang2015}, among others exotic behaviours.

The concept of strain engineering has been also extended to the optical context \cite{Bae2013,Ni2014,Chen2016,Rakheja2016}. The optical properties of graphene are ultimately provided by its electronic structure, which can be modified by strain. For example, pristine graphene presents a transparency defined by fundamental constants, around $97.7\%$, over a broad band of frequencies \cite{Nair2008}. This remarkable feature is essentially a consequence of its unusual low-energy electronic band structure around the Dirac points. Under uniform strain, such conical bands are deformed which produces anisotropy in the electronic dynamics \cite{Oliva13}. Accordingly, this effect gives rise an anisotropic optical conductivity of strained graphene \cite{Pellegrino2010,Pereira2010,Oliva2014,Oliva2014C} and, therefore, a modulation of its transmittance as a function of the polarization of the incident light, as experimentally observed \cite{Ni2014}. From a theoretical viewpoint, this optoelectronic behaviour of strained graphene has been quantified by continuum approaches up to first-order in the strain tensor \cite{Pereira2010,Oliva2014,Oliva2014C,Oliva2015}. However, nowadays there are novel methods for applying uniaxial strain larger than $10\%$ in a nondestructive and controlled manner \cite{Perez2014}. So, a low-energy continuum theory for the electronic and optical properties of strained graphene, up to second-order in the strain tensor, seems to be needed \cite{Ramezani2013,Li2014,Crosse2014,Shallcross2016}.

In this paper, we derive the effective Dirac Hamiltonian for graphene under uniform strain up to to second-order in the strain tensor. For this purpose, we start from a nearest-neighbor tight-binding model and carry out an expansion around the real Dirac point. Unlike previous approaches to the first-order in strain, we show how the obtained low-energy Hamiltonian reveals the trigonal symmetry of the graphene. Also, we calculate the optical conductivity of strained graphene and characterize its transmittance for a uniaxial strain up to second-order in the stretching magnitude.
These findings describe in a more accurate form the electronic and optical properties of strained graphene and, hence, can be potentially utilized towards novel optical characterizations of the strain state of graphene.

\section{Tight-binding model as starting point}

Strain effects on electronic properties of graphene are usually captured by using a nearest-neighbor tight-binding model \cite{Pereira09a,Zenan13,Salvador13,Mikkel16b}. Within this approach, one can demonstrate that the Hamiltonian in momentum space for graphene under a uniform strain is given by \cite{Pereira09a,Oliva13}
\begin{equation}\label{HT}
H(\bm{k})=-\sum_{n=1}^{3} t_{n} \left(
\begin{array}{cc}
0 & e^{-i\bm{k}\cdot\bm{\delta}_{n}^{\prime}}\\
e^{i\bm{k}\cdot\bm{\delta}_{n}^{\prime}} & 0
\end{array}\right),
\end{equation}
where the strained nearest-neighbor vectors are obtained by $\bm{\delta}_{n}^{\prime}=(\bar{\bm{I}}+\bar{\bm{\epsilon}})\cdot\bm{\delta}_{n}$, being $\bar{\bm{I}}$ the $(2\times2)$ identity matrix and $\bar{\bm{\epsilon}}$ the rank-two strain tensor, whose components are independent on the position. Here, we choose the unstrained nearest-neighbor vectors as
\begin{equation}
\bm{\delta}_{1}=\frac{a_{0}}{2}(\sqrt{3},1), \ \
\bm{\delta}_{2}=\frac{a_{0}}{2}(-\sqrt{3},1),\ \
\bm{\delta}_{3}=a_{0}(0,-1),
\end{equation}
where $a_{0}$ is the intercarbon distance for pristine graphene. Thus, the $x$ $(y)$ axis of the Cartesian coordinate system is along the zigzag (armchair) direction of the honeycomb lattice.
Owing to the changes in the intercabon distance, the nearest-neighbor hopping parameters are modified. Here we consider this effect by means of the commonly used model \cite{Papaconstantopoulos,Pereira09a,Ribeiro09,Mikkel16a}
\begin{equation}\label{tn}
t_{n}=t_{0}e^{-\beta(\vert\bm{\delta}_{n}^{\prime}\vert/a_{0}-1)},
\end{equation}
where $t_{0}=2.7\mbox{eV}$ is the hopping parameter for pristine graphene and $\beta\approx3$.

From equation (\ref{HT}) follows that the dispersion relation near the Fermi energy of graphene under uniform strain is given by two bands,
\begin{equation}\label{DR}
E(\bm{k})=\pm\vert t_{1}e^{i\bm{k}\cdot\bm{\delta}_{1}^{\prime}} + t_{2}e^{i\bm{k}\cdot\bm{\delta}_{2}^{\prime}} + t_{3}e^{i\bm{k}\cdot\bm{\delta}_{3}^{\prime}}\vert,
\end{equation}
which remains gapless as long as the triangular inequalities, $\vert t_{1}-t_{2}\vert\leq\vert t_{3}\vert\leq\vert t_{1}+t_{2}\vert$, are satisfied \cite{Hasegawa}. Evaluating equation (\ref{DR}) for uniaxial strains, V. Pereira, \emph{et al.}, found the minimum uniaxial deformation that leads to the gap opening is about $23\%$ \cite{Pereira09a}. This result is confirmed by the \emph{ab initio} calculations, finding that this gap in strained graphene requires deformations larger than $20\%$ \cite{Choi10,Ni09}. Therefore, the use of an effective Dirac Hamiltonian obtained from equation (\ref{HT}) is justified for uniform deformations up to the order of $10\%$.

For this purpose, it is important to take into account a crucial detail: \emph{the strain-induced shift of the Dirac points in momentum space}. In absence of deformation, the Dirac points $\bm{K}_{D}$ (determined by condition $E(\bm{K}_{D})=0$) coincide with the corners of the first Brillouin zone. Then, to obtain the effective Dirac Hamiltonian in this case, one simply expand the Hamiltonian (\ref{HT}) around such corners, e.g., $\bm{K}_{0}=(\frac{4\pi}{3\sqrt{3}a_{0}},0)$. However, in presence of deformations, the Dirac points do not coincide even with the corners of the strained first Brillouin zone \cite{Pereira09a,YangLi10}. Thus, to obtain the effective Dirac Hamiltonian, one should no longer expand the Hamiltonian (\ref{HT}) around $\bm{K}_{0}$. As demonstrated \cite{Oliva15b,Volovik14a}, such expansion around $\bm{K}_{0}$ yields an incorrect derivation of the anisotropic Fermi velocity. The appropriate procedure is to find first the new positions of the Dirac points and then carry out the expansion around them \cite{Yang2011,Oliva15b,Volovik14a,Bahat10,Volovik15}.

\section{Effective Dirac Hamiltonian}

As first step, we determine the new positions of Dirac points from the condition $E(\bm{K}_{D})=0$, up to second order in the strain tensor, which is the leading order used throughout the rest of the paper. Essentially, we calculate the strain-induced shift of the Dirac point $\bm{K}_{D}$ from the corner $\bm{K}_{0}$ of the first Brillouin zone by using equation $E(\bm{K}_{D})=0$, which leads to
\begin{equation}\label{E0}
 \sum_{n=1}^{3}t_{n}e^{i\bm{K}_{D}\cdot\bm{\delta}_{n}^{\prime}}=
 \sum_{n=1}^{3}t_{n}e^{i\bm{K}_{D}\cdot (\bar{\bm{I}}+\bar{\bm{\epsilon}})\cdot\bm{\delta}_{n}}=
 \sum_{n=1}^{3}t_{n}e^{i\bm{G}\cdot\bm{\delta}_{n}}=0,
\end{equation}
where $\bm{G}\equiv(\bar{\bm{I}}+\bar{\bm{\epsilon}})\cdot\bm{K}_{D}$ is the effective Dirac point. As demonstrated in Appendix~\ref{AA}, $\bm{G}$ can be expressed as
\begin{equation}\label{GD}
\bm{G}=\bm{K}_{0} + \bm{A}^{(1)} + \bm{A}^{(2)} + \mathcal{O}(\bar{\bm{\epsilon}}^{3}),
\end{equation}
where
\begin{eqnarray}
A_{x}^{(1)} + i A_{y}^{(1)}=\frac{\beta}{2a_{0}}(\epsilon_{xx}-\epsilon_{yy} - 2i\epsilon_{xy}),\label{A1}
\end{eqnarray}
and
\begin{eqnarray}
A_{x}^{(2)} + i A_{y}^{(2)}=\frac{\beta(4\beta+1)}{16a_{0}}(\epsilon_{xx}-\epsilon_{yy} + 2i\epsilon_{xy})^{2}.\label{A2}
\end{eqnarray}

\begin{figure}[b]
\includegraphics[width=8.5cm]{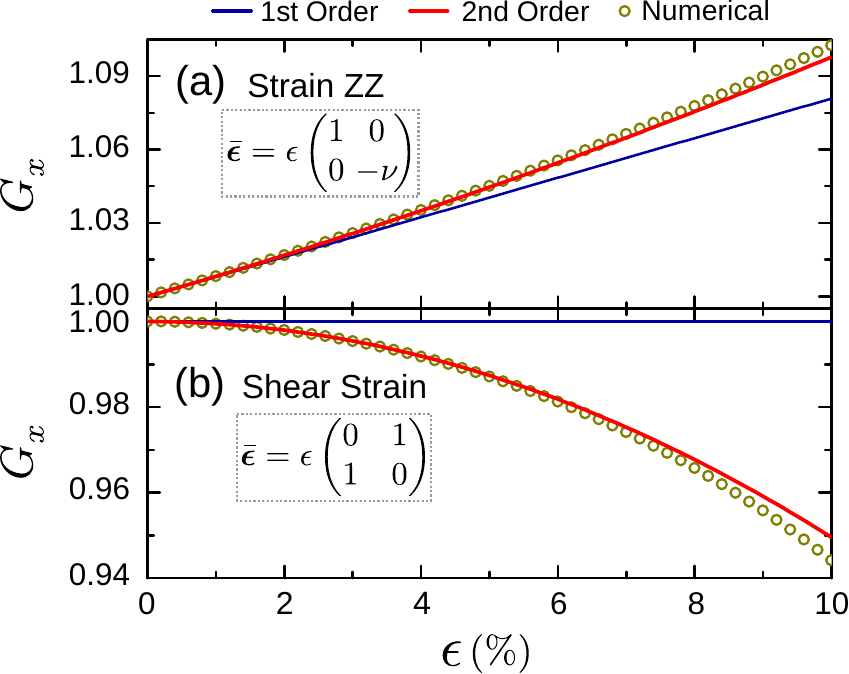}
\caption{The $x$-component of $\bm{G}$ ($G_{x}$) in units of $4\pi/(3\sqrt{3}a_{0})$ as a function of the strain magnitude $\epsilon$ for two different deformations. Panel (a) corresponds to a uniaxial strain along zigzag (ZZ) direction $(\epsilon_{xx}=\epsilon,\epsilon_{yy}=-\nu\epsilon,\epsilon_{xy}=0)$ and panel (b) corresponds to a shear strain $(\epsilon_{xx}=\epsilon_{yy}=0,\epsilon_{xy}=\epsilon)$. The blue and red lines are respectively the values of $G_{x}$ calculated up to first- and second-order in the strain tensor, while the open circles present the numerical values obtained from equation (\ref{E0}).}\label{fig1}
\end{figure}

Notice that the correction up to first order, $\bm{A}^{(1)}$, coincides with the value previously reported \cite{Oliva13}, which is interpreted as a gauge field for nonuniform deformations \cite{Yang2011,Oliva15b,Volovik14a}. On the other hand, the expression (\ref{A2}) for the second-order correction $\bm{A}^{(2)}$ is one of the main contributions of this work. To demonstrate its relevance, we numerically calculate the positions of $\bm{G}$ for two deformations and compare them with the analytical results given by (\ref{GD}-\ref{A2}). As illustrated in Fig.~\ref{fig1}(a) for a uniaxial strain along zigzag direction, the values of $G_{x}$ estimated up to first order in the strain magnitude $\epsilon$ (blue solid circles) clearly differ from the exact numerical values of $G_{x}$ (gray line) as $\epsilon$ increases, while the values of $G_{x}$ estimated up to second order (red open circles) show a significantly better approximation. The case of a shear strain is an even more illustrative example of the relevance of $\bm{A}^{(2)}$. According to the first-order correction, $G_{x}$ does not change under a shear strain, which is at variance with the exact numerical result displayed in Fig.~\ref{fig1}(b). In contrast, the values of $G_{x}$ estimated up to second order present a good agreement with the numerical values over the studied range of $\epsilon$. Beyond the present work, the second-order correction $\bm{A}^{(2)}$ for nonuniform strain could be relevant to a more complete analysis of the strain-induced pseudomagnetic fields. For example, in presence of a deformation field given by $\bm{u}=(u(y),0)$, for which $\bar{\epsilon}_{xx}=\bar{\epsilon}_{yy}=0$ and $\bar{\epsilon}_{xy}=\partial_{y}u(y)/2$, the pseudomagnetic field $B_{ps}$, derived from the standard expression $B_{ps}=\nabla\times\bm{A}^{(1)}$, results equal to zero. However, if $\bm{A}^{(2)}$ is taken into account by means of the possible generalized expression $B_{ps}=\nabla\times\bm{A}^{(1)} + \nabla\times\bm{A}^{(2)}$, one can demonstrate that the resulting pseudomagnetic field $B_{ps}$ is not zero. The implications of this issue will be discussed with details in an upcoming work.

Knowing the position of the Dirac point $\bm{K}_{D}$, through equation (\ref{GD}), one can now proceed to the expansion of Hamiltonian (\ref{HT}) around $\bm{K}_{D}$, by means of $\bm{k}=\bm{K}_{D}+\bm{q}$, to obtain the effective Dirac Hamiltonian. Following this approach up to second order in the strain tensor $\bar{\bm{\epsilon}}$, the effective Dirac Hamiltonian can be written as (see Appendix~\ref{AB})
\begin{equation}\label{DH}
H=\hbar v_{0} \bm{\tau}\cdot(\bar{\bm{I}} + \bar{\bm{\epsilon}} - \beta\bar{\bm{\epsilon}} -\beta\bar{\bm{\epsilon}}^{2} - \beta\bar{\bm{\kappa}}_{1} + \beta^{2}\bar{\bm{\kappa}}_{2})\cdot\bm{q},
\end{equation}
where $v_{0}=3t_{0}a_{0}/2\hbar$ is the Fermi velocity for pristine graphene, $\bm{\tau}=(\tau_{x},\tau_{y})$ is a vector of ($2\times2$) Pauli matrices describing the pseudospin degree of freedom,
\begin{equation}\label{kappa1}
\bar{\bm{\kappa}}_{1}=\frac{1}{8}
\left(\begin{array}{cc}
(\epsilon_{xx}-\epsilon_{yy})^{2} & -2\epsilon_{xy}(\epsilon_{xx}-\epsilon_{yy})\\
-2\epsilon_{xy}(\epsilon_{xx}-\epsilon_{yy}) & 4\epsilon_{xy}^{2}
\end{array}\right),
\end{equation}
and
\begin{equation}\label{kappa2}
\bar{\bm{\kappa}}_{2}=\frac{1}{4}
\left(\begin{array}{cc}
\epsilon_{xx}^{2}-\epsilon_{yy}^{2}+2\epsilon_{xx}\epsilon_{yy}+2\epsilon_{xy}^{2} & 4\epsilon_{xx}\epsilon_{xy}\\
4\epsilon_{xx}\epsilon_{xy} & 2(\epsilon_{yy}^{2}-\epsilon_{xy}^{2})
\end{array}\right).
\end{equation}

It is important to emphasize that the explicit form of equations (\ref{kappa1}) and (\ref{kappa2}) is a consequence of the Cartesian coordinate system $xy$ chosen. For an arbitrary coordinate system $\tilde{x}\tilde{y}$, rotated by an angle $\vartheta$ respect to the system $xy$, the new expressions for $\bar{\bm{\kappa}}_{1}$ and $\bar{\bm{\kappa}}_{2}$ should be found by means of the transformation rules of a second order Cartesian tensor \cite{Barber}.

From  equation (\ref{DH}) one can recognize the Fermi velocity tensor as
\begin{equation}\label{GV}
\bar{\bm{v}}=v_{0}(\bar{\bm{I}} + \bar{\bm{\epsilon}} - \beta\bar{\bm{\epsilon}} -\beta\bar{\bm{\epsilon}}^{2} - \beta\bar{\bm{\kappa}}_{1} + \beta^{2}\bar{\bm{\kappa}}_{2}),
\end{equation}
which generalizes the expression, $v_{0}(\bar{\bm{I}} + \bar{\bm{\epsilon}} - \beta\bar{\bm{\epsilon}})$, for the Fermi velocity tensor up to first-order in the strain tensor reported  in Refs. [\onlinecite{Oliva13,Volovik14a}].

As a consistency test, let us consider an isotropic uniform strain of the graphene lattice, which is simply given by $\bar{\bm{\epsilon}}=\epsilon\bar{\bm{I}}$. Under this deformation, the new intercarbon distance $a$ is rescaled as $a=a_{0}(1+\epsilon)$, whereas the new hopping parameter $t$, expanding equation (\ref{tn}) up to second order in strain, results $t=t_{0}(1-\beta\epsilon+\beta^{2}\epsilon^{2}/2)$. Therefore, the new Fermi velocity, $v=3t a/2\hbar$, obtained straight away from the nearest-neighbor tight-binding Hamiltonian, takes the value $v=v_{0}(1-\beta\epsilon+\epsilon-\beta\epsilon^{2}+\beta^{2}\epsilon^{2}/2)$. This result can be alternatively obtained by evaluating our tensor (\ref{GV}) for $\bar{\bm{\epsilon}}=\epsilon\bar{\bm{I}}$.

The tensorial character of $\bar{\bm{v}}$ is due to the elliptic shape of the isoenergetic curves around $\bm{K}_{D}$. Notice that the principal axes of the Fermi velocity tensor up to first-order in the strain tensor, $v_{0}(\bar{\bm{I}} + \bar{\bm{\epsilon}} - \beta\bar{\bm{\epsilon}})$, are collinear with the principal axes of $\bar{\bm{\epsilon}}$. Therefore, within the effective low-energy Hamiltonian up to first-order in the strain tensor, the anisotropic electronic behaviour is only originated from the strain-induced anisotropy. Nevertheless, the terms $\bar{\bm{\kappa}}_{1}$ and $\bar{\bm{\kappa}}_{2}$ in equation (\ref{GV}) suggest that the second-order deformation theory might reveal the anisotropy (trigonal symmetry) of the underlying honeycomb lattice.

\begin{figure}[b]
\includegraphics[width=7.5cm]{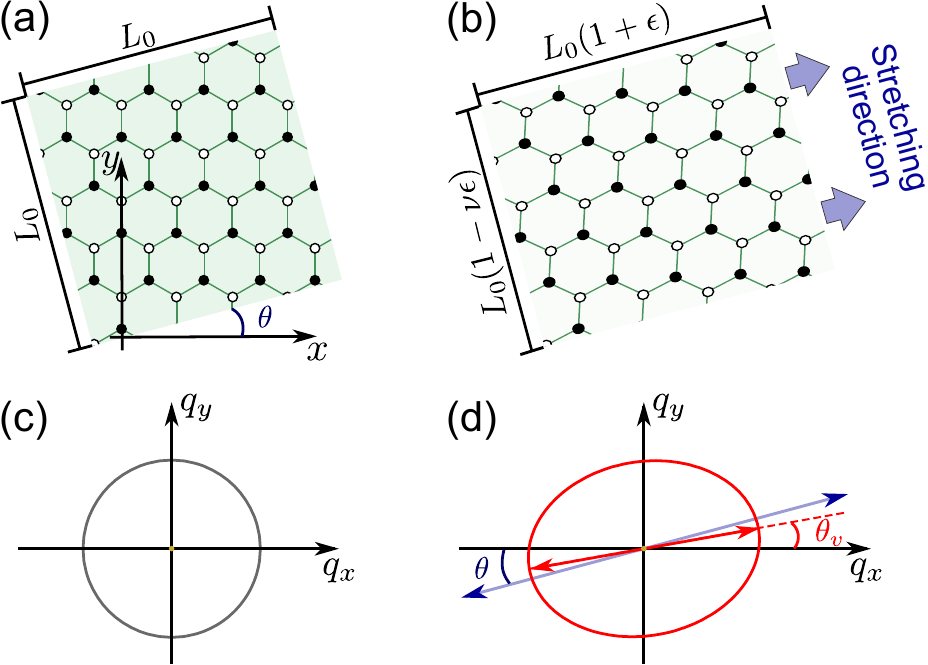}
\caption{Schematic representation of a portion of graphene (a) without and (b) with an applied uniaxial strain along an arbitrary angle $\theta$ respect to the zigzag direction ($x$-axis). Panels (c) and (d) illustrate the isoenergetic curves around Dirac ponits in the reciprocal space of graphene lattices at the deformation states (a) and (b), respectively, where $\theta_{v}$ determines the principal axis direction of the isoenergetic ellipse.}\label{fig2}
\end{figure}

To clarify this issue, let us to consider graphene subjected a uniaxial strain such that the stretching direction is rotated by an arbitrary angle $\theta$ respect to the Cartesian coordinate system $xy$ (see Fig.~\ref{fig2}). In this case, the strain tensor ($\bar{\bm{\epsilon}}$) in the reference system $xy$ reads
\begin{equation}\label{TD}
\bar{\bm{\epsilon}}(\theta)=\epsilon
\left(
\begin{array}{cc}
\cos^{2}\theta - \nu\sin^{2}\theta & (1+\nu)\cos\theta\sin\theta\\
(1+\nu)\cos\theta\sin\theta & \sin^{2}\theta - \nu\cos^{2}\theta
\end{array}
\right),
\end{equation}
where $\epsilon$ is the strain magnitude. Note that both $\bar{\bm{\epsilon}}(\theta)$ and $\bar{\bm{\epsilon}}(\theta+180^{\circ})$ represent physically the same uniaxial strain, which can be confirmed in equation (\ref{TD}). It is important to mention that for $\theta=n 60^{\circ}$ ($\theta=90^{\circ}+n 60^{\circ}$), being $n$ an integer, the stretching is along a zigzag (armchair) direction of graphene lattice.

As discussed above, under the strain (\ref{TD}), the Fermi velocity tensor up to first-order in the strain tensor, $v_{0}(\bar{\bm{I}} + \bar{\bm{\epsilon}} - \beta\bar{\bm{\epsilon}})$, is diagonal in the coordinate system $x'y'$, rotated by the angle $\theta$ respect to the coordinate system $xy$. However, the Fermi velocity tensor (\ref{GV}), up to second-order in the strain tensor, is diagonal in a coordinate system $x''y''$, rotated by an angle $\theta_{v}$ such that
\begin{equation}\label{tantheta}
\tan2\theta_{v}=\frac{2v_{xy}}{v_{xx}-v_{yy}},
\end{equation}
which determines the direction of lower electronic velocity. In the reciprocal space, the angle $\theta_{v}$ characterizes the pulling direction of isoenergetic curves, \emph{i.e.}, the principal axis of the isoenergetic ellipses, as illustrated in Fig.~\ref{fig2}(d).

\begin{figure}[b]
\includegraphics[width=8cm]{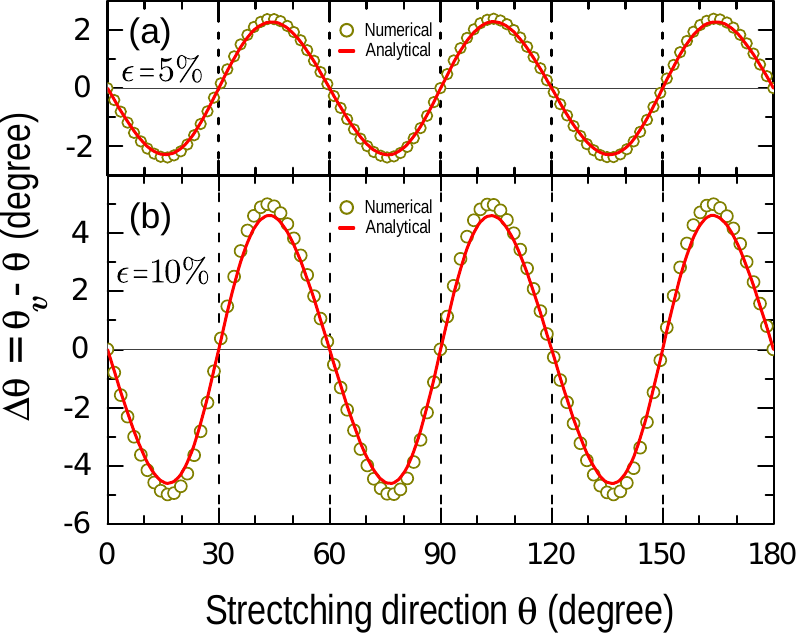}
\caption{Difference $\triangle \theta = \theta_{v}-\theta$ as a function of the stretching direction $\theta$ for two different strain magnitudes $\epsilon=5\%$ and $\epsilon=10\%$. Open circles correspond with the numerical values directly estimated from equation (\ref{tantheta}), while red lines are plotted by using the analytical expression (\ref{deltatheta}).}\label{fig3}
\end{figure}

In Fig. (\ref{fig3}), we show the difference $\triangle \theta = \theta_{v}-\theta$, numerically calculated from equation (\ref{tantheta}), as a function of the stretching direction $\theta$ for two different strain magnitudes $\epsilon=5\%$ and $10\%$. The observed six-fold behaviour of $\triangle \theta$ can be analytically evaluated by
\begin{eqnarray}\label{deltatheta}
\triangle \theta &\approx& -\frac{\beta(2\beta+1)(1+\nu)}{16(\beta-1)}\epsilon\sin(6\theta) \nonumber\\
&&\times\left(1-\frac{\beta(1-\nu)}{2}\epsilon\cos(6\theta)\right)
\end{eqnarray}
in good agreement with the numerical values, as shown in Fig. (\ref{fig3}). From the last expression, it follows that the principal axes of the Fermi velocity tensor (\ref{GV}) are only collinear with the principal axes of $\bar{\bm{\epsilon}}(\theta)$ for $\theta=n 30^{\circ}$, \emph{i.e.}, when the stretching is along the zigzag or armchair crystallographic directions. This result demonstrates that our Hamiltonian (\ref{DH}), a second-order deformation theory, reveals the trigonal symmetry of underlying honeycomb lattice.

\section{Optical properties}

An anisotropic Dirac system described by the effective Hamiltonian
\begin{equation}\label{DSmt}
H=\hbar
v_{0}\bm{\tau}\cdot(\bar{\bm{I}}+\bar{\bm{\Delta}})\cdot\bm{q},
\end{equation}
being $\bar{\bm{\Delta}}$ a symmetric ($2\times2$) matrix such that $\Delta_{ij}\ll1$, presents an anisotropic optical response captured by the conductivity tensor (see Appendix~\ref{AC}):
\begin{eqnarray}\label{CVFmt}
\bar{\bm{\sigma}}(\omega) &\approx& \sigma_{0}(\omega)\Bigl\{ \bar{\bm{I}} - \mbox{tr}(\bar{\bm{\Delta}})\bar{\bm{I}} + 2\bar{\bm{\Delta}} + \bar{\bm{\Delta}}^{2} \nonumber\\
&+& \frac{1}{2}\bigl[\bigl(\mbox{tr}(\bar{\bm{\Delta}})\bigr)^{2}
+ \mbox{tr}(\bar{\bm{\Delta}}^{2})\bigr]\bar{\bm{I}}-2\mbox{tr}(\bar{\bm{\Delta}})\bar{\bm{\Delta}} \Bigr\},
\end{eqnarray}
where $\omega$ is the frequency of the external electric field and $\sigma_{0}(\omega)$ is the optical conductivity of the unperturbed Dirac system, \emph{i.e.}, the optical conductivity of unstrained graphene. Equation (\ref{CVFmt}) is a generalization up to second-order in $\bar{\bm{\Delta}}$ of previous expression until first-order in $\bar{\bm{\Delta}}$ for the optical conductivity of an anisotropic Dirac system, as it can be seen in equation (17) of Ref.~[\onlinecite{Oliva2016}].

Now, comparing equations (\ref{DH}) and (\ref{DSmt}), the optical conductivity tensor $\bar{\bm{\sigma}}(\omega)$ of strained graphene is  straightforward obtained by making the replacement:
\begin{equation}
\bar{\bm{\Delta}}= \bar{\bm{\epsilon}} - \beta\bar{\bm{\epsilon}} -\beta\bar{\bm{\epsilon}}^{2} - \beta\bar{\bm{\kappa}}_{1} + \beta^{2}\bar{\bm{\kappa}}_{2},
\end{equation}
into equation (\ref{CVFmt}). Regarding terms up to second-order in the strain tensor, it results
\begin{eqnarray}\label{CondE}
\bar{\bm{\sigma}}(\omega)&=&\sigma_{0}(\omega) \Biggl[ \bar{\bm{I}} + \tilde{\beta}\mbox{tr}(\bar{\bm{\epsilon}})\bar{\bm{I}} - 2\tilde{\beta}\bar{\bm{\epsilon}} \nonumber\\
&+& \biggl(\frac{5\beta+2\tilde{\beta}^{2}}{4}\mbox{tr}(\bar{\bm{\epsilon}}^{2}) + \frac{4\tilde{\beta}^{2}-2\beta^{2}-\beta}{8}(\mbox{tr}(\bar{\bm{\epsilon}}))^{2}  \biggr)\bar{\bm{I}} \nonumber\\
&+& (\tilde{\beta}^{2}-2\beta)\bar{\bm{\epsilon}}^{2} -2\tilde{\beta}^{2}\mbox{tr}(\bar{\bm{\epsilon}})\bar{\bm{\epsilon}}  - 2\beta\bar{\bm{\kappa}}_{1} +2\beta^{2}\bar{\bm{\kappa}}_{2} \Biggr],
\end{eqnarray}
where $\tilde{\beta}=\beta-1$. This equation generalizes previous works \cite{Pereira2010,Pellegrino2010,Oliva2014,Oliva2014C}, in which the optical conductivity of graphene under uniform strain was reported up to first-order in the strain tensor. 

Let us make a proof about the consistency of equation (\ref{CondE}). When graphene is at half filling, \emph{i.e.}, the chemical potential
equals to zero, the optical conductivity $\sigma_{0}(\omega)$ is frequency-independent and is given by the universal value $e^{2}/(4\hbar)$ \cite{Ziegler07,Gusynin07}. It is important to emphasize that this result is independent on the value $v_{0}$ of the Fermi velocity \cite{Stauber2015}. Therefore, under an isotropic uniform strain $\bar{\bm{\epsilon}}=\epsilon\bar{\bm{I}}$, which only leads to a new isotropic Fermi velocity $v=v_{0}(1-\beta\epsilon+\epsilon-\beta\epsilon^{2}+\beta^{2}\epsilon^{2}/2)$, the optical conductivity does not change and remains equal to $\sigma_{0}=e^{2}/(4\hbar)$, at least within the Dirac cone approximation \cite{Stauber2015}. In other words, any expression reported as optical conductivity tensor for uniformly strained graphene, as a function on the strain tensor, to be evaluated for $\bar{\bm{\epsilon}}=\epsilon\bar{\bm{I}}$ must give rise $\sigma_{0}\bar{\bm{I}}$, as occurred when one evaluates the tensor (\ref{CondE}).

\begin{figure}[t]
\includegraphics[width=6.5cm]{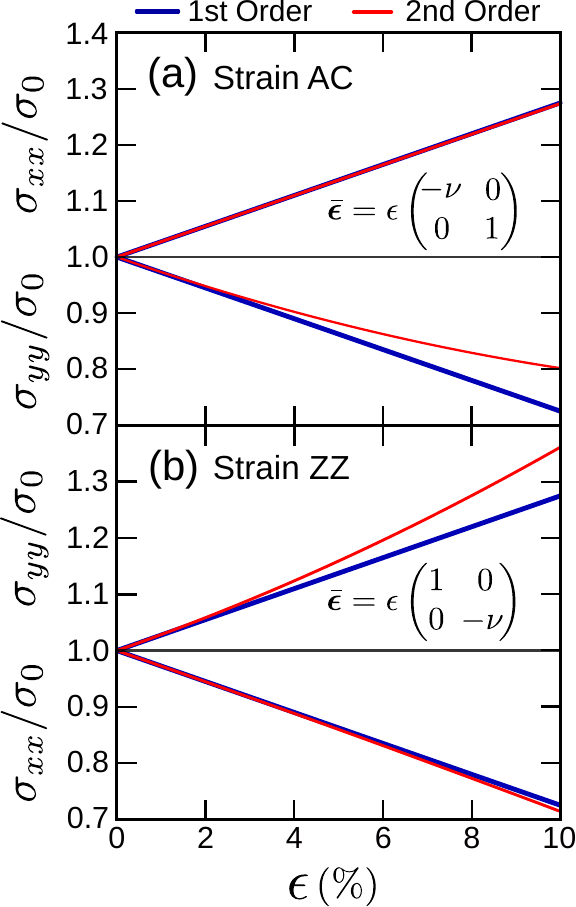}
\caption{Components, $\sigma_{xx}$ and $\sigma_{yy}$, of the optical conductivity tensor in units of $\sigma_{0}$ as functions of the strain magnitude $\epsilon$ for two different uniaxial deformations along the (a) armchair (AC) and (b) zigzag (ZZ) direction.}\label{fig4}
\end{figure}

The optical conductivity up to first-order in the strain tensor, $\sigma_{0}[\bar{\bm{I}} + \tilde{\beta}\mbox{tr}(\bar{\bm{\epsilon}})\bar{\bm{I}} - 2\tilde{\beta}\bar{\bm{\epsilon}}]$, under a uniaxial strain (\ref{TD}) can be characterized by $\sigma_{\parallel}=\sigma_{0}[1-\tilde{\beta}\epsilon(1+\nu)]$ and $\sigma_{\perp}=\sigma_{0}[1+\tilde{\beta}\epsilon(1+\nu)]$, where $\sigma_{\parallel}\ (\sigma_{\perp})$ is the optical conductivity parallel (perpendicular) to the stretching direction (see blue lines in Fig.~\ref{fig4}). Within the first-order approximation, the optical conductivity along the stretching direction decreases by the same amount that the transverse conductivity increases, independently of $\theta$. This behaviour is modified when second-order terms are taken into account.

In Fig.~\ref{fig4}(a), we plot the components of the optical conductivity tensor (\ref{CondE}) versus the stretching magnitude $\epsilon$ for a uniaxial strain along the armchair direction. The perpendicular conductivity to the stretching direction, $\sigma_{xx}(\epsilon)$, does not have appreciable difference respect to the lineal approximation $\sigma_{0}[1+\tilde{\beta}\epsilon(1+\nu)]$ whereas the parallel conductivity, $\sigma_{yy}(\epsilon)$, noticeably differs from $\sigma_{0}[1-\tilde{\beta}\epsilon(1+\nu)]$ with increasing strain. On the other hand, Fig.~\ref{fig4}(b) displays a contrary behaviour of the optical conductivity for a uniaxial strain along the zigzag direction. For this case, the parallel conductivity, $\sigma_{xx}(\epsilon)$, looks slight different from $\sigma_{0}[1-\tilde{\beta}\epsilon(1+\nu)]$ whereas the perpendicular conductivity, $\sigma_{yy}(\epsilon)$, is noticeably greater than $\sigma_{0}[1+\tilde{\beta}\epsilon(1+\nu)]$ with increasing strain. This increase of $\sigma_{yy}(\epsilon)$ respect to $\sigma_{0}[1+\tilde{\beta}\epsilon(1+\nu)]$ might help to give a better understanding of the change in the transmission of hybrid graphene integrated microfibers elongated along their axial direction \cite{Chen2016}. For example, in Figure 2(b) of Ref.~[\onlinecite{Chen2016}], it is possible to appreciate that the experimental data of this change gradually differ, with increasing strain, from the theoretical calculation using the first-order linear approximation $\sigma_{0}[1+\tilde{\beta}\epsilon(1+\nu)]$, which can be improved by considering the second-order contribution as shown in Fig. \ref{fig4}(b). 

To complete our discussion about the emergence of the trigonal symmetry of graphene in the continuum approach presented here, we now study the transmittance of linearly polarized light on strained graphene. Considering graphene as a two-dimensional sheet with conductivity $\bar{\bm{\sigma}}$ and from the boundary conditions, vacuum-graphene-vacuum, for the electromagnetic field on the interfaces, the transmittance for normal incidence reads as \cite{Stauber2008,Oliva2015}
\begin{equation}\label{T}
T=\left(1 + \frac{\mbox{Re}[\sigma_{xx}\cos^{2}\theta_{i} + \sigma_{yy}\sin^{2}\theta_{i} + \sigma_{xy}\sin 2\theta_{i}]}{2\varepsilon_{0}c}\right)^{-2}
\end{equation}
where $\varepsilon_{0}$ is the vacuum permittivity, $c$ is the speed of light in vacuum and $\theta_{i}$ is the incident polarization angle. Note that for a pristine graphene with $\bar{\bm{\sigma}}=\sigma_{0}\bar{\bm{I}}$, equation (\ref{CondE}) reproduces the experimentally observed constant transmittance $T=(1+\pi\alpha/2)^{-2}\approx1-\pi\alpha$ over visible and infrared spectrum \cite{Nair2008}, being $\alpha\approx1/137$ the fine-structure constant.
From equation (\ref{T}) it can be seen that an anisotropic absorbance yields a periodic modulation of the transmittance as a function of the polarization direction $\theta_{i}$ \cite{Pereira2010,Oliva2015,Ni2014}.

\begin{figure}[t]
\includegraphics[width=\linewidth]{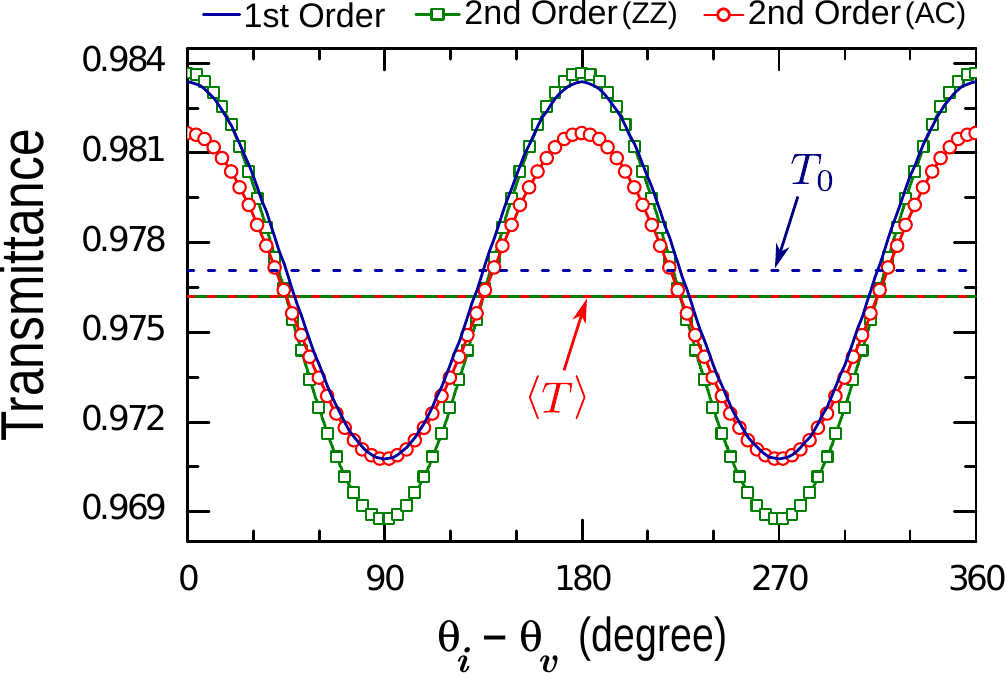}
\caption{Transmittance of equation (\ref{TUD}) as a function of the angular difference $\theta_{i}-\theta_{v}$ for two uniaxial strain with the same magnitude of $\epsilon=10\%$ but different stretching directions. The green circles and red squares respectively correspond to stretching along the zigzag (ZZ) and armchair (AC) directions. The blue line is the transmittance according to the first-order theory.}\label{fig5}
\end{figure}

For the case of a uniaxial strain, and assuming the chemical potential
equal to zero, from equations (\ref{TD}), (\ref{CondE}) and (\ref{T}) it follows that the transmittance up to second-order in the strain magnitude $\epsilon$ is given by
\begin{eqnarray}\label{TUD}
T &=& 1 - \pi\alpha + \pi\alpha\tilde{\beta}(1+\nu)\epsilon\cos2(\theta_{i}-\theta_{v}) - \frac{\pi\alpha}{2}\tilde{\beta}^{2}(1+\nu)^{2}\epsilon^{2} \nonumber\\
&& + \frac{\pi\alpha}{2}(1+\gamma\cos6\theta)(1+\nu)^{2}\epsilon^{2}\cos2(\theta_{i}-\theta_{v}),
\end{eqnarray}
where $\gamma=\beta(2\beta+1)/4$. Expression (\ref{TUD}) reveals two new remarkable features in comparison with the first-order theory. As illustrated in Fig.~\ref{fig5}, the transmittance mean value, $\langle T\rangle= 1 - \pi\alpha - \pi\alpha\tilde{\beta}^{2}(1+\nu)^{2}\epsilon^{2}/2$, has a negative shift with respect to the first-order average value $T_{0} = 1 - \pi\alpha$. Second, the transmittance oscillation amplitude ($\triangle T$) is determined by
\begin{equation}\label{AmpT}
\triangle T (\theta) = 2\pi\alpha\tilde{\beta}(1+\nu)\epsilon + \pi\alpha(1+\gamma\cos6\theta)(1+\nu)^{2}\epsilon^{2}.
\end{equation}

While the first-order expression for the transmittance oscillation amplitude, $2\pi\alpha\tilde{\beta}(1+\nu)\epsilon$, is independent on the stretching direction $\theta$, $\triangle T$ of equation (\ref{AmpT}) depends on $\theta$. For example, for a uniaxial strain along the zigzag (armchair) direction with $\theta=n 60^{\circ}$ ($\theta=90^{\circ} + n 60^{\circ}$), $\triangle T$ takes its highest (lowest) value, as displayed in Fig. \ref{fig5}. This strectching direction dependent $\triangle T$ might be used to confirm experimentally the present theory up to second-order in the strain tensor, as done for small strain less than $1\%$ \cite{Ni2014}.

\section{Conclusion}

We have analytically deduced a new effective Dirac Hamiltonian of graphene under a uniform deformation up to second-order in the strain tensor, including new Dirac-point positions that are qualitatively different from those predicted by first-order approaches, as occurred for the shear strain. Moreover, based on a detailed analysis about the anisotropic Fermi velocity tensor, we demonstrated how our second-order deformation theory reveals the trigonal symmetry of graphene unlike the previous first-order results.

We further derived, for the first time, analytical expressions for the high-frequency electric conductivity and light transmittance of a strained graphene up to second-order in the strain tensor. The magnitude of this transmittance oscillates according to the incident light polarization and the oscillation amplitude depends on the stretching direction, in contrast to the first-order prediction. In fact, within the first-order theory, the maximal transmittance occurs when the light polarization coincides to the stretching direction. However, the second-order theory predicts such coincidences only for stretching along zigzag and armchair directions. Therefore, the obtained light transmittance results can be experimentally verified by optical absorption measurements and they would be used for characterizing the deformation states of strained graphene. In general, the analytical study presented in this article has the advantage of being concise and establishes a reference point for upcoming numerical and experimental investigations.

It would be important to stress that the observed absence of lattice symmetry in the optical properties of strained graphene is due to the combination of the low-energy effective Dirac model and first-order approximation in the strain tensor. Such absence can be overcome by carrying out the study within the first-neighbour tight-binding model as occurred for high-energy electron excitations \cite{Pereira09a} or by introducing second-order effects in the strain tensor even within the simplest Dirac model, as done in this article. This finding of trigonal symmetry in optical response reveals the capability of low-energy effective Dirac theory to describe properly anisotropic electron behaviour in graphene under strong uniform deformations. However, a tight-binding model beyond nearest-neoghbour interactions would be required to analyze both the gap opening and the electron-hole spectrum symmetry induced by lattice strain \cite{Neto09}. Finally, the present work can be extended to perform an analytical study of the pseudomagnetic fields induced by nonuniform strains.

\begin{acknowledgments}
This work has been partially supported by CONACyT of Mexico through Project 252943, and by PAPIIT of Universidad Nacional Aut\'onoma de M\'exico (UNAM) through Projects IN113714 and IN106317. Computations were performed at Miztli of UNAM. M.O.L. acknowledges the postdoctoral fellowship from DGAPA-UNAM.
\end{acknowledgments}

\appendix

\section{Dirac point position}\label{AA}

Here we provide the derivation of expressions (\ref{GD}--\ref{A2}) of main text.
Equation, $E(\bm{K}_{D})=0$, can be rewritten as
\begin{equation}\label{E0A}
 \sum_{n=1}^{3}t_{n}e^{i\bm{K}_{D}\cdot\bm{\delta}_{n}^{\prime}}=
 \sum_{n=1}^{3}t_{n}e^{i\bm{K}_{D}\cdot (\bar{\bm{I}}+\bar{\bm{\epsilon}})\cdot\bm{\delta}_{n}}=
 \sum_{n=1}^{3}t_{n}e^{i\bm{G}\cdot\bm{\delta}_{n}}=0,
\end{equation}
where $\bm{G}\equiv(\bar{\bm{I}}+\bar{\bm{\epsilon}})\cdot\bm{K}_{D}$ is the effective Dirac point associated to a pristine honeycomb lattice with strained nearest-neighbor hopping integrals $t_{n}$. To solve equation (\ref{E0A}) in a perturbative manner, we cast the position of $\bm{G}$ as
\begin{equation}\label{ExpG}
 \bm{G}=\bm{K}_{0} + \bm{A}^{(1)} + \bm{A}^{(2)} + \mathcal{O}(\bar{\bm{\epsilon}}^{3}),
\end{equation}
where $\bm{A}^{(1)}$ ( $\bm{A}^{(2)}$) is the correction from first (second) order in the strain tensor. Similarly, we consider Taylor expansions of $t_{n}$, up to second order in strain tensor, in the form
\begin{equation}\label{Exptn}
 t_{n}=t_{0}[1 + \Delta_{n}^{(1)} +  \Delta_{n}^{(2)} + \mathcal{O}(\bar{\bm{\epsilon}}^{3})],
\end{equation}
where $\Delta_{n}^{(1)}$ ($\Delta_{n}^{(2)}$) are terms of the first (second) order in the strain tensor.

Substituting equations~(\ref{ExpG}) and (\ref{Exptn}) into equation~(\ref{E0A}), the coefficient of the first-order strain tensor should be equal to zero, which leads to
\begin{equation}\label{EcA1}
\sum_{n=1}^{3}[\Delta_{n}^{(1)}+ i\bm{A}^{(1)}\cdot\bm{\delta}_{n}]e^{i\bm{K}_{0}\cdot\bm{\delta}_{n}}=0.
\end{equation}

Analogously, the coefficient of the second-order strain tensor should also be zero, yields
\begin{eqnarray}\label{EcA2}
\sum_{n=1}^{3} &&\bigl[ \Delta_{n}^{(2)} + i\Delta_{n}^{(1)}\bm{A}^{(1)}\cdot\bm{\delta}_{n} - (\bm{A}^{(1)}\cdot\bm{\delta}_{n})^{2}/2\bigr. \nonumber\\
&& \bigl. + i\bm{A}^{(2)}\cdot\bm{\delta}_{n}\bigr] e^{i\bm{K}_{0}\cdot\bm{\delta}_{n}}=0.
\end{eqnarray}

From equation (\ref{EcA1}), $\bm{A}^{(1)}$ can be determined and it is used as input of equation (\ref{EcA2}) to obtain $\bm{A}^{(2)}$. To carry out this procedure, it is necessary to explicitly know $\Delta_{n}^{(1)}$ and $\Delta_{n}^{(2)}$  as functions of the strain tensor.

Expanding $t_{n}$, up to second order in strain tensor, gives
\begin{eqnarray}
t_{n}/t_{0} &=& \exp[-\beta(\vert\bm{\delta}_{n}^{\prime}\vert/a_{0}-1)]\nonumber \\
&=& \exp\left[-\beta\left( \frac{1}{a_{0}^{2}}\bm{\delta}_{n}\cdot\bar{\bm{\epsilon}}\cdot\bm{\delta}_{n} + \frac{1}{2a_{0}^{4}}(\bar{\bm{\epsilon}}\cdot\bm{\delta}_{n})^{2}\right.\right.\nonumber\\
&& \left.\left. - \frac{1}{2a_{0}^{4}}(\bm{\delta}_{n}\cdot\bar{\bm{\epsilon}}\cdot\bm{\delta}_{n})^{2} + \mathcal{O}(\bar{\bm{\epsilon}}^{3})\right)\right]\nonumber\\
&=& 1 - \frac{\beta}{a_{0}^{2}}\bm{\delta}_{n}\cdot\bar{\bm{\epsilon}}\cdot\bm{\delta}_{n} - \frac{\beta}{2a_{0}^{4}}(\bar{\bm{\epsilon}}\cdot\bm{\delta}_{n})^{2}\nonumber\\ 
&& + \frac{\beta(\beta+1)}{2a_{0}^{4}}(\bm{\delta}_{n}\cdot\bar{\bm{\epsilon}}\cdot\bm{\delta}_{n})^{2} + \mathcal{O}(\bar{\bm{\epsilon}}^{3}).\label{Ec3}
\end{eqnarray}
Then, by comparing equations (\ref{Exptn}) and (\ref{Ec3}) one obtains
\begin{equation}
\Delta_{n}^{(1)}= - \frac{\beta}{a_{0}^{2}}\bm{\delta}_{n}\cdot\bar{\bm{\epsilon}}\cdot\bm{\delta}_{n},
\end{equation}
and 
\begin{equation}
\Delta_{n}^{(2)}= - \frac{\beta}{2a_{0}^{4}}(\bar{\bm{\epsilon}}\cdot\bm{\delta}_{n})^{2} + \frac{\beta(\beta+1)}{2a_{0}^{4}}(\bm{\delta}_{n}\cdot\bar{\bm{\epsilon}}\cdot\bm{\delta}_{n})^{2}.
\end{equation}

Finally, substituting $\Delta_{n}^{(1)}$ into equation (\ref{EcA1}), we get
\begin{equation}\label{A1A}
A_{x}^{(1)} + i A_{y}^{(1)}=\frac{\beta}{2a_{0}}(\epsilon_{xx}-\epsilon_{yy} - 2i\epsilon_{xy}),
\end{equation}
and consequently, using this result and the expression of $\Delta_{n}^{(2)}$, equation (\ref{EcA2}) can be rewritten as
\begin{equation}\label{A2A}
A_{x}^{(2)} + i A_{y}^{(2)}=\frac{\beta(4\beta+1)}{16a_{0}}(\epsilon_{xx}-\epsilon_{yy} + 2i\epsilon_{xy})^{2}.
\end{equation}

Note that equations (\ref{A1A}) and (\ref{A2A}) are the first- and second-order corrections to the Dirac point position given in equation (\ref{E0}) of the main text.

\vspace{0.5cm}

\section{Effective Dirac Hamiltonian}\label{AB}

In order to derive the effective Dirac Hamiltonian given by equation (\ref{DH}) in the main text, we start from the tight-binding model in momentum space for graphene under a uniform strain,
\begin{equation}\label{HTB}
H=-\sum_{n=1}^{3} t_{n} \left(
\begin{array}{cc}
0 & e^{-i\bm{k}\cdot(\bm{I}+\bar{\bm{\epsilon}})\cdot\bm{\delta}_{n}}\\
e^{i\bm{k}\cdot(\bm{I}+\bar{\bm{\epsilon}})\cdot\bm{\delta}_{n}} & 0
\end{array}\right),
\end{equation}
and we consider momenta close to the Dirac point $\bm{K}_{D}$, by means of the substitution $\bm{k}=\bm{K}_{D}+\bm{q}$. Then, expression (\ref{HTB}) transforms as
\begin{equation}\label{HT2}
H= \left(
\begin{array}{cc}
0 & h^{\ast}\\
h & 0
\end{array}\right),
\end{equation}
where $h=-\sum_{n=1}^{3} t_{n} e^{i(\bm{K}_{D}+\bm{q})\cdot(\bm{I}+\bar{\bm{\epsilon}})\cdot\bm{\delta}_{n}}$.
Now, using equation (\ref{ExpG}), $h$ can be expanded up to first-order in $\bm{q}$ and second-order in $\bar{\bm{\epsilon}}$ as
\begin{widetext}
\begin{eqnarray}\label{h12}
h&=& -\sum_{n=1}^{3} t_{n} e^{i[\bm{K}_{D}\cdot(\bm{I}+\bar{\bm{\epsilon}})\cdot\bm{\delta}_{n}+ \bm{q}\cdot(\bm{I}+\bar{\bm{\epsilon}})\cdot\bm{\delta}_{n}]} \nonumber \\
&\approx&  -\sum_{n=1}^{3} t_{n} e^{i\bm{K}_{0}\cdot\bm{\delta}_{n}} e^{i(\bm{A}^{(1)}\cdot\bm{\delta}_{n} + \bm{A}^{(2)}\cdot\bm{\delta}_{n})}e^{i \bm{q}\cdot(\bm{I}+\bar{\bm{\epsilon}})\cdot\bm{\delta}_{n}} \nonumber \\
&\approx& -\sum_{n=1}^{3} t_{n} e^{i\bm{K}_{0}\cdot\bm{\delta}_{n}} \left[1 + i \bm{A}^{(1)}\cdot\bm{\delta}_{n} + i \bm{A}^{(2)}\cdot\bm{\delta}_{n} - (\bm{A}^{(1)}\cdot\bm{\delta}_{n})^{2}/2\right]\Bigl[1 + i\bm{q}\cdot(\bm{I}+\bar{\bm{\epsilon}})\cdot\bm{\delta}_{n}\Bigr] \nonumber \\
&\approx& -\sum_{n=1}^{3} t_{n} e^{i\bm{K}_{0}\cdot\bm{\delta}_{n}} \left[1 + i \bm{A}^{(1)}\cdot\bm{\delta}_{n} + i \bm{A}^{(2)}\cdot\bm{\delta}_{n} - (\bm{A}^{(1)}\cdot\bm{\delta}_{n})^{2}/2 + i\bm{q}\cdot(\bm{I}+\bar{\bm{\epsilon}})\cdot\bm{\delta}_{n}\right. \nonumber\\
&& \ \ \ \ \ \ \ \ \ \ \ \ \ \ \ \ \ \ \ \ \ - \left. (\bm{A}^{(1)}\cdot\bm{\delta}_{n})(\bm{q}\cdot(\bm{I}+\bar{\bm{\epsilon}})\cdot\bm{\delta}_{n}) - (\bm{A}^{(2)}\cdot\bm{\delta}_{n})(\bm{q}\cdot\bm{\delta}_{n}) - i(\bm{A}^{(1)}\cdot\bm{\delta}_{n})^{2}(\bm{q}\cdot\bm{\delta}_{n})/2 \right]
\end{eqnarray}
and substituting expression (\ref{Exptn}) for $t_{n}$ in equation (\ref{h12}), the expansion of $h$ results
\begin{eqnarray}
h&\approx& -t_{0}\sum_{n=1}^{3} e^{i\bm{K}_{0}\cdot\bm{\delta}_{n}} \Bigl[ \underbrace{1 + i \bm{A}^{(1)}\cdot\bm{\delta}_{n} + i \bm{A}^{(2)}\cdot\bm{\delta}_{n} - (\bm{A}^{(1)}\cdot\bm{\delta}_{n})^{2}/2 + \Delta_{n}^{(1)} + i\Delta_{n}^{(1)}\bm{A}^{(1)}\cdot\bm{\delta}_{n} +  \Delta_{n}^{(2)}}_{suming\ over\ n\ equal\ to\ zero}\Bigr. \nonumber\\
&&\underbrace{+ i\bm{q}\cdot\bm{\delta}_{n}}_{h_{0}} \underbrace{+ i\bm{q}\cdot\bar{\bm{\epsilon}}\cdot\bm{\delta}_{n}}_{h_{1,a}} \underbrace{- (\bm{A}^{(1)}\cdot\bm{\delta}_{n})(\bm{q}\cdot\bm{\delta}_{n}) + i\Delta_{n}^{(1)}\bm{q}\cdot\bm{\delta}_{n}}_{h_{1,b}} \underbrace{-(\bm{A}^{(1)}\cdot\bm{\delta}_{n})(\bm{q}\cdot\bar{\bm{\epsilon}}\cdot\bm{\delta}_{n}) + i\Delta_{n}^{(1)}(\bm{q}\cdot\bar{\bm{\epsilon}}\cdot\bm{\delta}_{n})}_{h_{2,a}} \nonumber\\
&&\Bigl. \underbrace{- i(\bm{A}^{(1)}\cdot\bm{\delta}_{n})^{2}(\bm{q}\cdot\bm{\delta}_{n})/2 - \Delta_{n}^{(1)}(\bm{A}^{(1)}\cdot\bm{\delta}_{n})(\bm{q}\cdot\bm{\delta}_{n}) - (\bm{A}^{(2)}\cdot\bm{\delta}_{n})(\bm{q}\cdot\bm{\delta}_{n}) + i\Delta_{n}^{(2)}(\bm{q}\cdot\bm{\delta}_{n})}_{h_{2,b}} \Bigr].
\end{eqnarray}

\end{widetext}

By taking into account equation (\ref{EcA2}) and $\sum_{n=1}^{3}e^{i\bm{K}_{0}\cdot\bm{\delta}_{n}}=0$, the $\bm{q}$-independent terms in the last expression are cancelled. Thus, $h$ can be rewritten as
\begin{equation}\label{h}
h=h_{0}+h_{1,a}+h_{1,b}+h_{2,a},h_{2,b},
\end{equation}
where
\begin{equation}\label{h0}
h_{0}= - t_{0} \sum_{n=1}^{3} e^{i\bm{K}_{0}\cdot\bm{\delta}_{n}}  \Bigl[i\bm{q}\cdot\bm{\delta}_{n}\Bigr]=\frac{3t_{0}a_{0}}{2}(q_{x}+iq_{y}),
\end{equation}
\begin{eqnarray}\label{h1a}
h_{1,a}&=& - t_{0} \sum_{n=1}^{3} e^{i\bm{K}_{0}\cdot\bm{\delta}_{n}}  \Bigl[ i\bm{q}\cdot\bar{\bm{\epsilon}}\cdot\bm{\delta}_{n}\Bigr]\nonumber\\
&=& - t_{0} \sum_{n=1}^{3} e^{i\bm{K}_{0}\cdot\bm{\delta}_{n}}  \Bigl[ i\bm{Q}\cdot\bm{\delta}_{n}\Bigr]\nonumber\\
&=&\frac{3t_{0}a_{0}}{2}(Q_{x}+iQ_{y})\nonumber\\
&=&\frac{3t_{0}a_{0}}{2}\Bigl[\epsilon_{xx}q_{x} + \epsilon_{xy}q_{y} + i(\epsilon_{xy}q_{x} + \epsilon_{yy}q_{y})\Bigr],
\end{eqnarray}
\begin{eqnarray}\label{h1b}
h_{1,b}&=& - t_{0} \sum_{n=1}^{3} e^{i\bm{K}_{0}\cdot\bm{\delta}_{n}} (\bm{q}\cdot\bm{\delta}_{n}) \left[- (\bm{A}^{(1)}\cdot\bm{\delta}_{n}) + i\Delta_{n}^{(1)}\right]\nonumber\\
&=&-\frac{3t_{0}a_{0}}{2}\beta\Bigl[\epsilon_{xx}q_{x} + \epsilon_{xy}q_{y} + i(\epsilon_{xy}q_{x} + \epsilon_{yy}q_{y})\Bigr],
\end{eqnarray}

\begin{eqnarray}\label{h2a}
h_{2,a}&=& - t_{0} \sum_{n=1}^{3} e^{i\bm{K}_{0}\cdot\bm{\delta}_{n}} (\bm{q}\cdot\bar{\bm{\epsilon}}\cdot\bm{\delta}_{n}) \left[  -(\bm{A}^{(1)}\cdot\bm{\delta}_{n}) + i\Delta_{n}^{(1)} \right]\nonumber\\
&=& - t_{0} \sum_{n=1}^{3} e^{i\bm{K}_{0}\cdot\bm{\delta}_{n}} (\bm{Q}\cdot\bm{\delta}_{n})\left[  -(\bm{A}^{(1)}\cdot\bm{\delta}_{n}) + i\Delta_{n}^{(1)} \right]\nonumber\\
&=&-\frac{3t_{0}a_{0}}{2}\beta\Bigl[\epsilon_{xx}Q_{x} + \epsilon_{xy}Q_{y} + i(\epsilon_{xy}Q_{x} + \epsilon_{yy}Q_{y})\Bigr]\nonumber\\
&=& -\frac{3t_{0}a_{0}}{2}\beta\Bigl\{(\epsilon_{xx}^{2}+\epsilon_{xy}^{2})q_{x} + \epsilon_{xy}(\epsilon_{xx}+\epsilon_{xy})q_{y}\nonumber\\
&& + i\bigl[\epsilon_{xy}(\epsilon_{xx}+\epsilon_{xy})q_{x} + (\epsilon_{xx}^{2}+\epsilon_{xy}^{2})q_{y}\bigr]\Bigr\}
\end{eqnarray}
and
\begin{widetext}

\begin{eqnarray}\label{h2b}
h_{2,b}&=& - t_{0} \sum_{n=1}^{3} e^{i\bm{K}_{0}\cdot\bm{\delta}_{n}} (\bm{q}\cdot\bm{\delta}_{n}) \left[ - i(\bm{A}^{(1)}\cdot\bm{\delta}_{n})^{2}/2 - \Delta_{n}^{(1)}(\bm{A}^{(1)}\cdot\bm{\delta}_{n}) - (\bm{A}^{(2)}\cdot\bm{\delta}_{n}) + i\Delta_{n}^{(2)} \right]\nonumber\\
&=& -\frac{3t_{0}a_{0}}{2}\beta\frac{1}{8}\Bigl\{ (\epsilon_{xx}-\epsilon_{yy})^{2}q_{x} + 2\epsilon_{xy}(\epsilon_{xx}-\epsilon_{yy})q_{y} + i \bigl[2\epsilon_{xy}(\epsilon_{xx}-\epsilon_{yy})q_{x} - 4\epsilon_{xy}^{2}q_{y}\bigr] \Bigr\}\nonumber\\
&& + \frac{3t_{0}a_{0}}{2}\beta^{2}\frac{1}{4}\Bigl\{ (\epsilon_{xx}^{2}-\epsilon_{yy}^{2}+2\epsilon_{xx}\epsilon_{yy}+2\epsilon_{xy}^{2})q_{x} +  4\epsilon_{xx}\epsilon_{xy}q_{y} + i \bigl[4\epsilon_{xx}\epsilon_{xy}q_{x} + 2(\epsilon_{yy}^{2}-\epsilon_{xy}^{2})q_{y}\bigr] \Bigr\},
\end{eqnarray}
\end{widetext}
being $\bm{Q}=\bm{q}\cdot\bar{\bm{\epsilon}}$. To simplify each term of $h$ in equation (\ref{h}), we have used of $\bm{K}_{0}=(\frac{4\pi}{3\sqrt{3}a_{0}},0)$, equations (2) of the main text for $\bm{\delta}_{n}$ and the expressions obtained in the previous section for $\bm{A}^{(1)}, \bm{A}^{(2)}, \bm{\Delta}^{(1)}$ and $\bm{\Delta}^{(2)}$. In addition, note the same algebraic form between the initial expression of equation (\ref{h1a}) and equation(\ref{h0}) if one defines  $\bm{Q}=\bm{q}\cdot\bar{\bm{\epsilon}}$. This similarity is also observed between equations (\ref{h2a}) and (\ref{h1b}).

\vspace{0.5cm}

In consequence, using equations (\ref{h0}--\ref{h2b}) we obtain the contribution of each term of $h$ to equation (\ref{HT2}) as
\begin{equation}\label{Mh0}
\left(
\begin{array}{cc}
0 & h_{0}^{\ast}\\
h_{0} & 0
\end{array}\right)=
\hbar v_{0} \bm{\tau}\cdot\bm{q},
\end{equation}
\begin{equation}\label{Mh1a}
\left(
\begin{array}{cc}
0 & h_{1,a}^{\ast}\\
h_{1,a} & 0
\end{array}\right)=\hbar v_{0} \bm{\tau}\cdot\bar{\bm{\epsilon}}\cdot\bm{q},
\end{equation}
\vspace{-0.2cm}
\begin{equation}\label{Mh1b}
\left(
\begin{array}{cc}
0 & h_{1,b}^{\ast}\\
h_{1,b} & 0
\end{array}\right)=-\hbar v_{0} \beta\bm{\tau}\cdot\bar{\bm{\epsilon}}\cdot\bm{q},
\end{equation}
\begin{equation}\label{Mh2a}
\left(
\begin{array}{cc}
0 & h_{2,a}^{\ast}\\
h_{2,a} & 0
\end{array}\right)=-\hbar v_{0} \beta\bm{\tau}\cdot\bar{\bm{\epsilon}}^{2}\cdot\bm{q},
\end{equation}
\vspace{-0.2cm}
and
\vspace{-0.1cm}
\begin{equation}\label{Mh2b}
\left(
\begin{array}{cc}
0 & h_{2,b}^{\ast}\\
h_{2,b} & 0
\end{array}\right)=\hbar v_{0} \bm{\tau}\cdot(- \beta\bar{\bm{\kappa}}_{1} + \beta^{2}\bar{\bm{\kappa}}_{2})\cdot\bm{q},
\end{equation}
where $v_{0}=3t_{0}a_{0}/2\hbar$ is the Fermi velocity for pristine graphene, $\bm{\tau}=(\tau_{x},\tau_{y})$ is a vector of ($2\times2$) Pauli matrices, $\bar{\bm{\kappa}}_{1}$ and $\bar{\bm{\kappa}}_{2}$ are respectively the matrices (10) and (11) of the main text. To obtain the expressions (\ref{Mh0}--\ref{Mh2b}) in terms of Pauli matrices, we used the identity
\begin{widetext}
\begin{equation}\label{MEj}
\left(\begin{array}{cc}
0 & \chi_{xx}q_{x}+\chi_{xy}q_{y} - i[\chi_{xy}q_{x}+\chi_{yy}q_{y}]\\
\chi_{xx}q_{x}+\chi_{xy}q_{y} + i[\chi_{xy}q_{x}+\chi_{yy}q_{y}] & 0
\end{array}\right)=
\bm{\tau}\cdot\bar{\bm{\chi}}\cdot\bm{q},
\end{equation}
\end{widetext}
where $\chi_{ij}$ are the elements of an arbitrary ($2\times2$) matrix $\bar{\bm{\chi}}$, being $\bar{\bm{\chi}}=\bm{I}$, $\bar{\bm{\chi}}=\bar{\bm{\epsilon}}$,  $\bar{\bm{\chi}}=-\beta\bar{\bm{\epsilon}}$,  $\bar{\bm{\chi}}=-\beta\bar{\bm{\epsilon}}^{2}$ and $\bar{\bm{\chi}}=-\beta\bar{\bm{\kappa}}_{1} + \beta^{2}\bar{\bm{\kappa}}_{2}$ for equations (\ref{Mh0}) to (\ref{Mh2b}), respectively. It is worth mentioning that expression (\ref{Mh0}) is the effective Dirac Hamiltonian for pristine graphene, while (\ref{Mh1a}) and (\ref{Mh1b}) are corrections to first-order in $\bar{\bm{\epsilon}}$, previously derived in Ref.~[\onlinecite{Oliva13}]. The second-order corrections in $\bar{\bm{\epsilon}}$, equations (\ref{Mh2a}) and (\ref{Mh2b}), are among the principal contributions of our work.

Finally, combining equations (\ref{HT2}), (\ref{h}) and (\ref{Mh0}--\ref{Mh2b}), we obtain the effective Dirac Hamiltonian for graphene under a uniaxial strain, up to second-order in the strain tensor $\bar{\bm{\epsilon}}$, given by
\begin{equation}
H=\hbar v_{0} \bm{\tau}\cdot(\bar{\bm{I}} + \bar{\bm{\epsilon}} - \beta\bar{\bm{\epsilon}} -\beta\bar{\bm{\epsilon}}^{2} - \beta\bar{\bm{\kappa}}_{1} + \beta^{2}\bar{\bm{\kappa}}_{2})\cdot\bm{q},
\end{equation}
which is the equation (\ref{DH}) reported in the main text.

\section{Optical conductivity of an anisotropic Dirac system}\label{AC}

In this section, we derive the optical conductivity tensor $\bar{\sigma}_{ij}(\omega)$ of an anisotropic Dirac system described by the effective Hamiltonian
\begin{equation}\label{DS}
H=\hbar
v_{0}\bm{\tau}\cdot(\bar{\bm{I}}+\bar{\bm{\Delta}})\cdot\bm{q},
\end{equation}
where the anisotropic behaviour is expressed through the perturbation $\bar{\bm{\Delta}}$, which is a symmetric ($2\times2$) matrix such that $\bar{\Delta}_{ij}\ll1$. Essentially, we now extend, up to second-order in $\bar{\bm{\Delta}}$, a previous calculation of $\bar{\sigma}_{ij}(\omega)$ up to first-order in $\bar{\bm{\Delta}}$ reported in Ref. [\onlinecite{Oliva2016}].

Assuming that the considered system has linear response to an external electric
field of frequency $\omega$, its optical conductivity $\bar{\sigma}_{ij}(w)$ can be calculated by combining the Hamiltonian (\ref{DS}) and the Kubo formula. Following the approach used in
Refs.~[\onlinecite{Ziegler06,Ziegler07}], $\bar{\sigma}_{ij}(\omega)$ can be expressed as a double integral with respect to two energies $E$, $E'$:
\begin{eqnarray} \label{C}
\bar{\sigma}_{ij}(\omega) &=& i\frac{e^2}{\hbar}\int\int \text{Tr}\{ v_{i}\delta(H-E')v_{j} \delta(H-E)\} \nonumber\\
&&\times\frac{1}{E-E'+\hbar\omega - i\alpha}\frac{f(E)-f(E')}{E-E'}\text{d}
E\text{d} E',
\end{eqnarray}
where $f(E)=(1+\exp[E/(k_{B}T)])^{-1}$ is the Fermi function at
temperature $T$, $\mbox{Tr}$ is the trace operator including the summation over the $\bm{q}$-space (as defined in equation (7) of Ref.~[\onlinecite{Ziegler07}]) and $v_{l}=i [H,r_{l}]$ is the velocity operator in the $l$-direction, with $l=x,y$.

To calculate the integral (\ref{C}) it is convenient to make the change of variables
\begin{equation}\label{CV}
\bm{q}=(\bar{\bm{I}}+\bar{\bm{\Delta}})^{-1}\cdot\bm{q}^{*},
\end{equation}
which yields that the Hamiltonian (\ref{DS}) becomes
$H=\hbar v_{0}\bm{\tau}\cdot\bm{q}^{*}$, corresponding to the case of a unperturbed and isotropic Dirac system, as unstrained graphene. At the same time, the velocity operator components transform as
\begin{eqnarray}\label{jx}
v_{x}&=&i [H,r_{x}]=\frac{\partial H}{\partial q_{x}}, \nonumber\\
&=&\left(\frac{\partial H}{\partial q_{x}^{*}} \frac{\partial
q_{x}^{*}}{\partial q_{x}} + \frac{\partial H}{\partial q_{y}^{*}}
\frac{\partial q_{y}^{*}}{\partial q_{x}}\right), \nonumber\\
&=&(1 + \bar{\Delta}_{xx})v_{x}^{*}
+ \bar{\Delta}_{xy}v_{y}^{*},
\end{eqnarray}
and analogously
\begin{equation}\label{jy}
v_{y}=(1 + \bar{\Delta}_{yy})v_{y}^{*}
+ \bar{\Delta}_{xy}v_{x}^{*},
\end{equation}
where $v_{x}^{*}=(\partial H/\partial q_{x}^{*})$ and
$v_{y}^{*}=(\partial H/\partial q_{y}^{*})$ are the velocity
operator components for the unperturbed Dirac system.

Then, substituting equations (\ref{jx}) and (\ref{jy}) into
equation~(\ref{C}) we find
\begin{eqnarray}
\bar{\sigma}_{xx}(\omega)&=&\bigl[(1+\bar{\Delta}_{xx})^{2}+\bar{\Delta}_{xy}^{2}\bigr]J\sigma_{0}(\omega),\label{sxx}\\
\bar{\sigma}_{yy}(\omega)&=&\bigl[(1+\bar{\Delta}_{yy})^{2}+\bar{\Delta}_{xy}^{2}\bigr]J\sigma_{0}(\omega),
\end{eqnarray}
and
\begin{eqnarray}
\bar{\sigma}_{xy}(\omega)&=&\bar{\sigma}_{yx}(\omega)\nonumber\\
&=& \bigl[2\bar{\Delta}_{xy}+\bar{\Delta}_{xy}(\bar{\Delta}_{xx}+\bar{\Delta}_{yy})\bigr]J\sigma_{0}(\omega),\label{sxy}
\end{eqnarray}
where $J$ is the Jacobian determinant of the transformation (\ref{CV}) originated by expressing the trace operator $\mbox{Tr}$ of equation (\ref{C}) in the new variables $\bm{q}^{*}$ and $\sigma_{0}(\omega)$
is the optical conductivity of the unperturbed Dirac system, \emph{i.e.}, the reported optical conductivity of unstrained graphene \cite{Ziegler07,Gusynin07,Stauber2008}. Note that equations (\ref{sxx}-\ref{sxy}) can be written in a compact manner as
\begin{equation}\label{CondJ}
\bar{\bm{\sigma}}(\omega)=\bigr(\bar{\bm{I}} + 2\bar{\bm{\Delta}} + \bar{\bm{\Delta}}^{2}\bigr)J\sigma_{0}(\omega).
\end{equation}

Now, if $J$ is expressed up to second-order in $\bar{\bm{\Delta}}$, results
\begin{eqnarray}\label{J}
J&=&\mbox{det}\bigr[(\bar{\bm{I}}+\bar{\bm{\Delta}})^{-1}\bigl]
\approx\mbox{det}\bigr(\bar{\bm{I}}-\bar{\bm{\Delta}}+\bar{\bm{\Delta}}^{2}\bigl)\nonumber\\
&\approx& 1-\mbox{tr}(\bar{\bm{\Delta}}) + \bigr[\mbox{tr}(\bar{\bm{\Delta}})\bigl]^{2}-\mbox{det}(\bar{\bm{\Delta}})\nonumber\\
&\approx& 1-\mbox{tr}(\bar{\bm{\Delta}}) + \bigl[\mbox{tr}(\bar{\bm{\Delta}})\bigr]^{2}/2 + \mbox{tr}(\bar{\bm{\Delta}}^{2})/2,
\end{eqnarray}
where $\mbox{tr}(\bar{\bm{\Delta}})=\bar{\Delta}_{xx}+\bar{\Delta}_{yy}$.

Finally, substituting equation (\ref{J}) into (\ref{CondJ}) we obtain that the optical conductivity tensor of the anisotropic Dirac system, described by the Hamiltonian (\ref{DS}), is given by
\begin{eqnarray}\label{CVF}
\bar{\bm{\sigma}}(\omega) &\approx& \sigma_{0}(\omega)\Bigl\{ \bar{\bm{I}} - \mbox{tr}(\bar{\bm{\Delta}})\bar{\bm{I}} + 2\bar{\bm{\Delta}} + \bar{\bm{\Delta}}^{2} \nonumber\\
&+& \frac{1}{2}\bigl[\bigl(\mbox{tr}(\bar{\bm{\Delta}})\bigr)^{2}
+ \mbox{tr}(\bar{\bm{\Delta}}^{2})\bigr]\bar{\bm{I}}-2\mbox{tr}(\bar{\bm{\Delta}})\bar{\bm{\Delta}} \Bigr\},
\end{eqnarray}
where the second term is the contribution to second-order in the perturbation $\bar{\bm{\Delta}}$ to the conductivity.

\bibliography{biblioStrainedGraphene}

\begin{thebibliography}{61}%
\makeatletter
\providecommand \@ifxundefined [1]{%
 \@ifx{#1\undefined}
}%
\providecommand \@ifnum [1]{%
 \ifnum #1\expandafter \@firstoftwo
 \else \expandafter \@secondoftwo
 \fi
}%
\providecommand \@ifx [1]{%
 \ifx #1\expandafter \@firstoftwo
 \else \expandafter \@secondoftwo
 \fi
}%
\providecommand \natexlab [1]{#1}%
\providecommand \enquote  [1]{``#1''}%
\providecommand \bibnamefont  [1]{#1}%
\providecommand \bibfnamefont [1]{#1}%
\providecommand \citenamefont [1]{#1}%
\providecommand \href@noop [0]{\@secondoftwo}%
\providecommand \href [0]{\begingroup \@sanitize@url \@href}%
\providecommand \@href[1]{\@@startlink{#1}\@@href}%
\providecommand \@@href[1]{\endgroup#1\@@endlink}%
\providecommand \@sanitize@url [0]{\catcode `\\12\catcode `\$12\catcode
  `\&12\catcode `\#12\catcode `\^12\catcode `\_12\catcode `\%12\relax}%
\providecommand \@@startlink[1]{}%
\providecommand \@@endlink[0]{}%
\providecommand \url  [0]{\begingroup\@sanitize@url \@url }%
\providecommand \@url [1]{\endgroup\@href {#1}{\urlprefix }}%
\providecommand \urlprefix  [0]{URL }%
\providecommand \Eprint [0]{\href }%
\providecommand \doibase [0]{http://dx.doi.org/}%
\providecommand \selectlanguage [0]{\@gobble}%
\providecommand \bibinfo  [0]{\@secondoftwo}%
\providecommand \bibfield  [0]{\@secondoftwo}%
\providecommand \translation [1]{[#1]}%
\providecommand \BibitemOpen [0]{}%
\providecommand \bibitemStop [0]{}%
\providecommand \bibitemNoStop [0]{.\EOS\space}%
\providecommand \EOS [0]{\spacefactor3000\relax}%
\providecommand \BibitemShut  [1]{\csname bibitem#1\endcsname}%
\let\auto@bib@innerbib\@empty
\bibitem [{\citenamefont {Galiotis}\ \emph {et~al.}(2015)\citenamefont
  {Galiotis}, \citenamefont {Frank}, \citenamefont {Koukaras},\ and\
  \citenamefont {Sfyris}}]{Galiotis2015}%
  \BibitemOpen
  \bibfield  {author} {\bibinfo {author} {\bibfnamefont {Costas}\ \bibnamefont
  {Galiotis}}, \bibinfo {author} {\bibfnamefont {Otakar}\ \bibnamefont
  {Frank}}, \bibinfo {author} {\bibfnamefont {Emmanuel~N.}\ \bibnamefont
  {Koukaras}}, \ and\ \bibinfo {author} {\bibfnamefont {Dimitris}\ \bibnamefont
  {Sfyris}},\ }\bibfield  {title} {\enquote {\bibinfo {title} {{Graphene
  Mechanics: Current Status and Perspectives}},}\ }\href {\doibase
  10.1146/annurev-chembioeng-061114-123216} {\bibfield  {journal} {\bibinfo
  {journal} {Annual Review of Chemical and Biomolecular Engineering}\ }\textbf
  {\bibinfo {volume} {6}},\ \bibinfo {pages} {121--140} (\bibinfo {year}
  {2015})}\BibitemShut {NoStop}%
\bibitem [{\citenamefont {Daniels}\ \emph {et~al.}(2015)\citenamefont
  {Daniels}, \citenamefont {Horning}, \citenamefont {Phillips}, \citenamefont
  {Massote}, \citenamefont {Liang}, \citenamefont {Bullard}, \citenamefont
  {Sumpter},\ and\ \citenamefont {Meunier}}]{Colin2015}%
  \BibitemOpen
  \bibfield  {author} {\bibinfo {author} {\bibfnamefont {Colin}\ \bibnamefont
  {Daniels}}, \bibinfo {author} {\bibfnamefont {Andrew}\ \bibnamefont
  {Horning}}, \bibinfo {author} {\bibfnamefont {Anthony}\ \bibnamefont
  {Phillips}}, \bibinfo {author} {\bibfnamefont {Daniel V~P}\ \bibnamefont
  {Massote}}, \bibinfo {author} {\bibfnamefont {Liangbo}\ \bibnamefont
  {Liang}}, \bibinfo {author} {\bibfnamefont {Zachary}\ \bibnamefont
  {Bullard}}, \bibinfo {author} {\bibfnamefont {Bobby~G}\ \bibnamefont
  {Sumpter}}, \ and\ \bibinfo {author} {\bibfnamefont {Vincent}\ \bibnamefont
  {Meunier}},\ }\bibfield  {title} {\enquote {\bibinfo {title} {{Elastic,
  plastic, and fracture mechanisms in graphene materials}},}\ }\href
  {http://stacks.iop.org/0953-8984/27/i=37/a=373002} {\bibfield  {journal}
  {\bibinfo  {journal} {Journal of Physics: Condensed Matter}\ }\textbf
  {\bibinfo {volume} {27}},\ \bibinfo {pages} {373002} (\bibinfo {year}
  {2015})}\BibitemShut {NoStop}%
\bibitem [{\citenamefont {Bissett}\ \emph {et~al.}(2014)\citenamefont
  {Bissett}, \citenamefont {Tsuji},\ and\ \citenamefont {Ago}}]{Bissett2014}%
  \BibitemOpen
  \bibfield  {author} {\bibinfo {author} {\bibfnamefont {Mark~A.}\ \bibnamefont
  {Bissett}}, \bibinfo {author} {\bibfnamefont {Masaharu}\ \bibnamefont
  {Tsuji}}, \ and\ \bibinfo {author} {\bibfnamefont {Hiroki}\ \bibnamefont
  {Ago}},\ }\bibfield  {title} {\enquote {\bibinfo {title} {{Strain engineering
  the properties of graphene and other two-dimensional crystals}},}\ }\href
  {\doibase 10.1039/C3CP55443K} {\bibfield  {journal} {\bibinfo  {journal}
  {Phys. Chem. Chem. Phys.}\ }\textbf {\bibinfo {volume} {16}},\ \bibinfo
  {pages} {11124--11138} (\bibinfo {year} {2014})}\BibitemShut {NoStop}%
\bibitem [{\citenamefont {Zhang}\ and\ \citenamefont
  {Zhang}(2015)}]{Zhang2015}%
  \BibitemOpen
  \bibfield  {author} {\bibinfo {author} {\bibfnamefont {Gang}\ \bibnamefont
  {Zhang}}\ and\ \bibinfo {author} {\bibfnamefont {Yong-Wei}\ \bibnamefont
  {Zhang}},\ }\bibfield  {title} {\enquote {\bibinfo {title} {Strain effects on
  thermoelectric properties of two-dimensional materials},}\ }\href {\doibase
  http://dx.doi.org/10.1016/j.mechmat.2015.03.009} {\bibfield  {journal}
  {\bibinfo  {journal} {Mechanics of Materials}\ }\textbf {\bibinfo {volume}
  {91, Part 2}},\ \bibinfo {pages} {382 -- 398} (\bibinfo {year}
  {2015})}\BibitemShut {NoStop}%
\bibitem [{\citenamefont {Si}\ \emph {et~al.}(2016)\citenamefont {Si},
  \citenamefont {Sun},\ and\ \citenamefont {Liu}}]{Si2016}%
  \BibitemOpen
  \bibfield  {author} {\bibinfo {author} {\bibfnamefont {Chen}\ \bibnamefont
  {Si}}, \bibinfo {author} {\bibfnamefont {Zhimei}\ \bibnamefont {Sun}}, \ and\
  \bibinfo {author} {\bibfnamefont {Feng}\ \bibnamefont {Liu}},\ }\bibfield
  {title} {\enquote {\bibinfo {title} {Strain engineering of graphene: a
  review},}\ }\href {\doibase 10.1039/C5NR07755A} {\bibfield  {journal}
  {\bibinfo  {journal} {Nanoscale}\ }\textbf {\bibinfo {volume} {8}},\ \bibinfo
  {pages} {3207--3217} (\bibinfo {year} {2016})}\BibitemShut {NoStop}%
\bibitem [{\citenamefont {Amorim}\ \emph {et~al.}(2016)\citenamefont {Amorim},
  \citenamefont {Cortijo}, \citenamefont {de~Juan}, \citenamefont {Grushin},
  \citenamefont {Guinea}, \citenamefont {Guti\'errez-Rubio}, \citenamefont
  {Ochoa}, \citenamefont {Parente}, \citenamefont {Rold\'an}, \citenamefont
  {San-Jose}, \citenamefont {Schiefele}, \citenamefont {Sturla},\ and\
  \citenamefont {Vozmediano}}]{Amorim2016}%
  \BibitemOpen
  \bibfield  {author} {\bibinfo {author} {\bibfnamefont {B.}~\bibnamefont
  {Amorim}}, \bibinfo {author} {\bibfnamefont {A.}~\bibnamefont {Cortijo}},
  \bibinfo {author} {\bibfnamefont {F.}~\bibnamefont {de~Juan}}, \bibinfo
  {author} {\bibfnamefont {A.G.}\ \bibnamefont {Grushin}}, \bibinfo {author}
  {\bibfnamefont {F.}~\bibnamefont {Guinea}}, \bibinfo {author} {\bibfnamefont
  {A.}~\bibnamefont {Guti\'errez-Rubio}}, \bibinfo {author} {\bibfnamefont
  {H.}~\bibnamefont {Ochoa}}, \bibinfo {author} {\bibfnamefont
  {V.}~\bibnamefont {Parente}}, \bibinfo {author} {\bibfnamefont
  {R.}~\bibnamefont {Rold\'an}}, \bibinfo {author} {\bibfnamefont
  {P.}~\bibnamefont {San-Jose}}, \bibinfo {author} {\bibfnamefont
  {J.}~\bibnamefont {Schiefele}}, \bibinfo {author} {\bibfnamefont
  {M.}~\bibnamefont {Sturla}}, \ and\ \bibinfo {author} {\bibfnamefont
  {M.A.H.}\ \bibnamefont {Vozmediano}},\ }\bibfield  {title} {\enquote
  {\bibinfo {title} {Novel effects of strains in graphene and other two
  dimensional materials},}\ }\href {\doibase
  http://dx.doi.org/10.1016/j.physrep.2015.12.006} {\bibfield  {journal}
  {\bibinfo  {journal} {Physics Reports}\ }\textbf {\bibinfo {volume} {617}},\
  \bibinfo {pages} {1 -- 54} (\bibinfo {year} {2016})}\BibitemShut {NoStop}%
\bibitem [{\citenamefont {Naumis}\ \emph {et~al.}(2016)\citenamefont {Naumis},
  \citenamefont {Barraza-Lopez}, \citenamefont {Oliva-Leyva},\ and\
  \citenamefont {Terrones}}]{Oliva2017}%
  \BibitemOpen
  \bibfield  {author} {\bibinfo {author} {\bibfnamefont {Gerardo~G.}\
  \bibnamefont {Naumis}}, \bibinfo {author} {\bibfnamefont {Salvador}\
  \bibnamefont {Barraza-Lopez}}, \bibinfo {author} {\bibfnamefont {Maurice}\
  \bibnamefont {Oliva-Leyva}}, \ and\ \bibinfo {author} {\bibfnamefont
  {Humberto}\ \bibnamefont {Terrones}},\ }\bibfield  {title} {\enquote
  {\bibinfo {title} {A review of the electronic and optical properties of
  strained graphene and other similar 2d materials},}\ }\href
  {https://arxiv.org/abs/1611.08627} {\bibfield  {journal} {\bibinfo  {journal}
  {arXiv:1611.08627}\ } (\bibinfo {year} {2016})},\ \bibinfo {note} {to be
  published in \emph{Rep. Prog. Phys.}}\BibitemShut {Stop}%
\bibitem [{\citenamefont {Pereira}\ \emph {et~al.}(2009)\citenamefont
  {Pereira}, \citenamefont {Castro~Neto},\ and\ \citenamefont
  {Peres}}]{Pereira09a}%
  \BibitemOpen
  \bibfield  {author} {\bibinfo {author} {\bibfnamefont {Vitor~M.}\
  \bibnamefont {Pereira}}, \bibinfo {author} {\bibfnamefont {A.~H.}\
  \bibnamefont {Castro~Neto}}, \ and\ \bibinfo {author} {\bibfnamefont
  {N.~M.~R.}\ \bibnamefont {Peres}},\ }\bibfield  {title} {\enquote {\bibinfo
  {title} {Tight-binding approach to uniaxial strain in graphene},}\ }\href
  {\doibase 10.1103/PhysRevB.80.045401} {\bibfield  {journal} {\bibinfo
  {journal} {Phys. Rev. B}\ }\textbf {\bibinfo {volume} {80}},\ \bibinfo
  {pages} {045401} (\bibinfo {year} {2009})}\BibitemShut {NoStop}%
\bibitem [{\citenamefont {Choi}\ \emph
  {et~al.}(2010{\natexlab{a}})\citenamefont {Choi}, \citenamefont {Jhi},\ and\
  \citenamefont {Son}}]{Choi2010}%
  \BibitemOpen
  \bibfield  {author} {\bibinfo {author} {\bibfnamefont {Seon-Myeong}\
  \bibnamefont {Choi}}, \bibinfo {author} {\bibfnamefont {Seung-Hoon}\
  \bibnamefont {Jhi}}, \ and\ \bibinfo {author} {\bibfnamefont {Young-Woo}\
  \bibnamefont {Son}},\ }\bibfield  {title} {\enquote {\bibinfo {title}
  {Effects of strain on electronic properties of graphene},}\ }\href {\doibase
  10.1103/PhysRevB.81.081407} {\bibfield  {journal} {\bibinfo  {journal} {Phys.
  Rev. B}\ }\textbf {\bibinfo {volume} {81}},\ \bibinfo {pages} {081407}
  (\bibinfo {year} {2010}{\natexlab{a}})}\BibitemShut {NoStop}%
\bibitem [{\citenamefont {Hasegawa}\ \emph {et~al.}(2006)\citenamefont
  {Hasegawa}, \citenamefont {Konno}, \citenamefont {Nakano},\ and\
  \citenamefont {Kohmoto}}]{Hasegawa}%
  \BibitemOpen
  \bibfield  {author} {\bibinfo {author} {\bibfnamefont {Yasumasa}\
  \bibnamefont {Hasegawa}}, \bibinfo {author} {\bibfnamefont {Rikio}\
  \bibnamefont {Konno}}, \bibinfo {author} {\bibfnamefont {Hiroki}\
  \bibnamefont {Nakano}}, \ and\ \bibinfo {author} {\bibfnamefont {Mahito}\
  \bibnamefont {Kohmoto}},\ }\bibfield  {title} {\enquote {\bibinfo {title}
  {Zero modes of tight-binding electrons on the honeycomb lattice},}\ }\href
  {\doibase 10.1103/PhysRevB.74.033413} {\bibfield  {journal} {\bibinfo
  {journal} {Phys. Rev. B}\ }\textbf {\bibinfo {volume} {74}},\ \bibinfo
  {pages} {033413} (\bibinfo {year} {2006})}\BibitemShut {NoStop}%
\bibitem [{\citenamefont {Guinea}\ \emph
  {et~al.}(2010{\natexlab{a}})\citenamefont {Guinea}, \citenamefont
  {Katsnelson},\ and\ \citenamefont {Geim}}]{Guinea2010a}%
  \BibitemOpen
  \bibfield  {author} {\bibinfo {author} {\bibfnamefont {F.}~\bibnamefont
  {Guinea}}, \bibinfo {author} {\bibfnamefont {M.~I.}\ \bibnamefont
  {Katsnelson}}, \ and\ \bibinfo {author} {\bibfnamefont {A.~K.}\ \bibnamefont
  {Geim}},\ }\bibfield  {title} {\enquote {\bibinfo {title} {Energy gaps and a
  zero-field quantum hall effect in graphene by strain engineering},}\ }\href
  {\doibase 10.1038/nphys1420} {\bibfield  {journal} {\bibinfo  {journal} {Nat
  Phys}\ }\textbf {\bibinfo {volume} {6}},\ \bibinfo {pages} {30--33} (\bibinfo
  {year} {2010}{\natexlab{a}})}\BibitemShut {NoStop}%
\bibitem [{\citenamefont {Guinea}\ \emph
  {et~al.}(2010{\natexlab{b}})\citenamefont {Guinea}, \citenamefont {Geim},
  \citenamefont {Katsnelson},\ and\ \citenamefont {Novoselov}}]{Guinea2010b}%
  \BibitemOpen
  \bibfield  {author} {\bibinfo {author} {\bibfnamefont {F.}~\bibnamefont
  {Guinea}}, \bibinfo {author} {\bibfnamefont {A.~K.}\ \bibnamefont {Geim}},
  \bibinfo {author} {\bibfnamefont {M.~I.}\ \bibnamefont {Katsnelson}}, \ and\
  \bibinfo {author} {\bibfnamefont {K.~S.}\ \bibnamefont {Novoselov}},\
  }\bibfield  {title} {\enquote {\bibinfo {title} {Generating quantizing
  pseudomagnetic fields by bending graphene ribbons},}\ }\href {\doibase
  10.1103/PhysRevB.81.035408} {\bibfield  {journal} {\bibinfo  {journal} {Phys.
  Rev. B}\ }\textbf {\bibinfo {volume} {81}},\ \bibinfo {pages} {035408}
  (\bibinfo {year} {2010}{\natexlab{b}})}\BibitemShut {NoStop}%
\bibitem [{\citenamefont {Gradinar}\ \emph {et~al.}(2013)\citenamefont
  {Gradinar}, \citenamefont {Mucha-Kruczy\ifmmode~\acute{n}\else
  \'{n}\fi{}ski}, \citenamefont {Schomerus},\ and\ \citenamefont
  {Fal'ko}}]{Falko2013}%
  \BibitemOpen
  \bibfield  {author} {\bibinfo {author} {\bibfnamefont {Diana~A.}\
  \bibnamefont {Gradinar}}, \bibinfo {author} {\bibfnamefont {Marcin}\
  \bibnamefont {Mucha-Kruczy\ifmmode~\acute{n}\else \'{n}\fi{}ski}}, \bibinfo
  {author} {\bibfnamefont {Henning}\ \bibnamefont {Schomerus}}, \ and\ \bibinfo
  {author} {\bibfnamefont {Vladimir~I.}\ \bibnamefont {Fal'ko}},\ }\bibfield
  {title} {\enquote {\bibinfo {title} {Transport signatures of pseudomagnetic
  landau levels in strained graphene ribbons},}\ }\href {\doibase
  10.1103/PhysRevLett.110.266801} {\bibfield  {journal} {\bibinfo  {journal}
  {Phys. Rev. Lett.}\ }\textbf {\bibinfo {volume} {110}},\ \bibinfo {pages}
  {266801} (\bibinfo {year} {2013})}\BibitemShut {NoStop}%
\bibitem [{\citenamefont {Bahamon}\ \emph {et~al.}(2015)\citenamefont
  {Bahamon}, \citenamefont {Qi}, \citenamefont {Park}, \citenamefont
  {Pereira},\ and\ \citenamefont {Campbell}}]{Bahamon2015}%
  \BibitemOpen
  \bibfield  {author} {\bibinfo {author} {\bibfnamefont {Dario~A.}\
  \bibnamefont {Bahamon}}, \bibinfo {author} {\bibfnamefont {Zenan}\
  \bibnamefont {Qi}}, \bibinfo {author} {\bibfnamefont {Harold~S.}\
  \bibnamefont {Park}}, \bibinfo {author} {\bibfnamefont {Vitor~M.}\
  \bibnamefont {Pereira}}, \ and\ \bibinfo {author} {\bibfnamefont {David~K.}\
  \bibnamefont {Campbell}},\ }\bibfield  {title} {\enquote {\bibinfo {title}
  {Conductance signatures of electron confinement induced by strained
  nanobubbles in graphene},}\ }\href {\doibase 10.1039/C5NR03393D} {\bibfield
  {journal} {\bibinfo  {journal} {Nanoscale}\ }\textbf {\bibinfo {volume}
  {7}},\ \bibinfo {pages} {15300--15309} (\bibinfo {year} {2015})}\BibitemShut
  {NoStop}%
\bibitem [{\citenamefont {Burgos}\ \emph {et~al.}(2015)\citenamefont {Burgos},
  \citenamefont {Warnes}, \citenamefont {Lima},\ and\ \citenamefont
  {Lewenkopf}}]{Burgos2015}%
  \BibitemOpen
  \bibfield  {author} {\bibinfo {author} {\bibfnamefont {Rhonald}\ \bibnamefont
  {Burgos}}, \bibinfo {author} {\bibfnamefont {Jesus}\ \bibnamefont {Warnes}},
  \bibinfo {author} {\bibfnamefont {Leandro R.~F.}\ \bibnamefont {Lima}}, \
  and\ \bibinfo {author} {\bibfnamefont {Caio}\ \bibnamefont {Lewenkopf}},\
  }\bibfield  {title} {\enquote {\bibinfo {title} {Effects of a random gauge
  field on the conductivity of graphene sheets with disordered ripples},}\
  }\href {\doibase 10.1103/PhysRevB.91.115403} {\bibfield  {journal} {\bibinfo
  {journal} {Phys. Rev. B}\ }\textbf {\bibinfo {volume} {91}},\ \bibinfo
  {pages} {115403} (\bibinfo {year} {2015})}\BibitemShut {NoStop}%
\bibitem [{\citenamefont {Settnes}\ \emph
  {et~al.}(2016{\natexlab{a}})\citenamefont {Settnes}, \citenamefont {Power},\
  and\ \citenamefont {Jauho}}]{Mikkel16a}%
  \BibitemOpen
  \bibfield  {author} {\bibinfo {author} {\bibfnamefont {Mikkel}\ \bibnamefont
  {Settnes}}, \bibinfo {author} {\bibfnamefont {Stephen~R.}\ \bibnamefont
  {Power}}, \ and\ \bibinfo {author} {\bibfnamefont {Antti-Pekka}\ \bibnamefont
  {Jauho}},\ }\bibfield  {title} {\enquote {\bibinfo {title} {Pseudomagnetic
  fields and triaxial strain in graphene},}\ }\href {\doibase
  10.1103/PhysRevB.93.035456} {\bibfield  {journal} {\bibinfo  {journal} {Phys.
  Rev. B}\ }\textbf {\bibinfo {volume} {93}},\ \bibinfo {pages} {035456}
  (\bibinfo {year} {2016}{\natexlab{a}})}\BibitemShut {NoStop}%
\bibitem [{\citenamefont {Stegmann}\ and\ \citenamefont
  {Szpak}(2016)}]{Stegmann2016}%
  \BibitemOpen
  \bibfield  {author} {\bibinfo {author} {\bibfnamefont {Thomas}\ \bibnamefont
  {Stegmann}}\ and\ \bibinfo {author} {\bibfnamefont {Nikodem}\ \bibnamefont
  {Szpak}},\ }\bibfield  {title} {\enquote {\bibinfo {title} {Current flow
  paths in deformed graphene: from quantum transport to classical trajectories
  in curved space},}\ }\href {http://stacks.iop.org/1367-2630/18/i=5/a=053016}
  {\bibfield  {journal} {\bibinfo  {journal} {New Journal of Physics}\ }\textbf
  {\bibinfo {volume} {18}},\ \bibinfo {pages} {053016} (\bibinfo {year}
  {2016})}\BibitemShut {NoStop}%
\bibitem [{\citenamefont {Settnes}\ \emph
  {et~al.}(2016{\natexlab{b}})\citenamefont {Settnes}, \citenamefont {Leconte},
  \citenamefont {Barrios-Vargas}, \citenamefont {Jauho},\ and\ \citenamefont
  {Roche}}]{Mikkel16b}%
  \BibitemOpen
  \bibfield  {author} {\bibinfo {author} {\bibfnamefont {Mikkel}\ \bibnamefont
  {Settnes}}, \bibinfo {author} {\bibfnamefont {Nicolas}\ \bibnamefont
  {Leconte}}, \bibinfo {author} {\bibfnamefont {Jose~E}\ \bibnamefont
  {Barrios-Vargas}}, \bibinfo {author} {\bibfnamefont {Antti-Pekka}\
  \bibnamefont {Jauho}}, \ and\ \bibinfo {author} {\bibfnamefont {Stephan}\
  \bibnamefont {Roche}},\ }\bibfield  {title} {\enquote {\bibinfo {title}
  {Quantum transport in graphene in presence of strain-induced pseudo-landau
  levels},}\ }\href {http://stacks.iop.org/2053-1583/3/i=3/a=034005} {\bibfield
   {journal} {\bibinfo  {journal} {2D Materials}\ }\textbf {\bibinfo {volume}
  {3}},\ \bibinfo {pages} {034005} (\bibinfo {year}
  {2016}{\natexlab{b}})}\BibitemShut {NoStop}%
\bibitem [{\citenamefont {Carrillo-Bastos}\ \emph {et~al.}(2016)\citenamefont
  {Carrillo-Bastos}, \citenamefont {Le\'on}, \citenamefont {Faria},
  \citenamefont {Latg\'e}, \citenamefont {Andrei},\ and\ \citenamefont
  {Sandler}}]{Carrillo2016}%
  \BibitemOpen
  \bibfield  {author} {\bibinfo {author} {\bibfnamefont {R.}~\bibnamefont
  {Carrillo-Bastos}}, \bibinfo {author} {\bibfnamefont {C.}~\bibnamefont
  {Le\'on}}, \bibinfo {author} {\bibfnamefont {D.}~\bibnamefont {Faria}},
  \bibinfo {author} {\bibfnamefont {A.}~\bibnamefont {Latg\'e}}, \bibinfo
  {author} {\bibfnamefont {E.~Y.}\ \bibnamefont {Andrei}}, \ and\ \bibinfo
  {author} {\bibfnamefont {N.}~\bibnamefont {Sandler}},\ }\bibfield  {title}
  {\enquote {\bibinfo {title} {Strained fold-assisted transport in graphene
  systems},}\ }\href {\doibase 10.1103/PhysRevB.94.125422} {\bibfield
  {journal} {\bibinfo  {journal} {Phys. Rev. B}\ }\textbf {\bibinfo {volume}
  {94}},\ \bibinfo {pages} {125422} (\bibinfo {year} {2016})}\BibitemShut
  {NoStop}%
\bibitem [{\citenamefont {Georgi~et al.}(2016)}]{Georgi2016}%
  \BibitemOpen
  \bibfield  {author} {\bibinfo {author} {\bibfnamefont {Alexander}\
  \bibnamefont {Georgi~et al.}},\ }\bibfield  {title} {\enquote {\bibinfo
  {title} {Tunable pseudo-zeeman effect in graphene},}\ }\href
  {https://arxiv.org/abs/1611.06123} {\bibfield  {journal} {\bibinfo  {journal}
  {arXiv:1611.06123}\ } (\bibinfo {year} {2016})}\BibitemShut {NoStop}%
\bibitem [{\citenamefont {de~Gail}\ \emph {et~al.}(2012)\citenamefont
  {de~Gail}, \citenamefont {Fuchs}, \citenamefont {Goerbig}, \citenamefont
  {Pi\'echon},\ and\ \citenamefont {Montambaux}}]{deGail2012}%
  \BibitemOpen
  \bibfield  {author} {\bibinfo {author} {\bibfnamefont {R.}~\bibnamefont
  {de~Gail}}, \bibinfo {author} {\bibfnamefont {J.-N.}\ \bibnamefont {Fuchs}},
  \bibinfo {author} {\bibfnamefont {M.O.}\ \bibnamefont {Goerbig}}, \bibinfo
  {author} {\bibfnamefont {F.}~\bibnamefont {Pi\'echon}}, \ and\ \bibinfo
  {author} {\bibfnamefont {G.}~\bibnamefont {Montambaux}},\ }\bibfield  {title}
  {\enquote {\bibinfo {title} {Manipulation of dirac points in graphene-like
  crystals},}\ }\href {\doibase http://dx.doi.org/10.1016/j.physb.2012.01.072}
  {\bibfield  {journal} {\bibinfo  {journal} {Physica B: Condensed Matter}\
  }\textbf {\bibinfo {volume} {407}},\ \bibinfo {pages} {1948 -- 1952}
  (\bibinfo {year} {2012})}\BibitemShut {NoStop}%
\bibitem [{\citenamefont {Naumis}\ and\ \citenamefont
  {Roman-Taboada}(2014)}]{Naumis2014}%
  \BibitemOpen
  \bibfield  {author} {\bibinfo {author} {\bibfnamefont {Gerardo~G.}\
  \bibnamefont {Naumis}}\ and\ \bibinfo {author} {\bibfnamefont {Pedro}\
  \bibnamefont {Roman-Taboada}},\ }\bibfield  {title} {\enquote {\bibinfo
  {title} {Mapping of strained graphene into one-dimensional hamiltonians:
  Quasicrystals and modulated crystals},}\ }\href {\doibase
  10.1103/PhysRevB.89.241404} {\bibfield  {journal} {\bibinfo  {journal} {Phys.
  Rev. B}\ }\textbf {\bibinfo {volume} {89}},\ \bibinfo {pages} {241404}
  (\bibinfo {year} {2014})}\BibitemShut {NoStop}%
\bibitem [{\citenamefont {Kauppila}\ \emph {et~al.}(2016)\citenamefont
  {Kauppila}, \citenamefont {Aikebaier},\ and\ \citenamefont
  {Heikkil\"a}}]{Kauppila2016}%
  \BibitemOpen
  \bibfield  {author} {\bibinfo {author} {\bibfnamefont {V.~J.}\ \bibnamefont
  {Kauppila}}, \bibinfo {author} {\bibfnamefont {F.}~\bibnamefont {Aikebaier}},
  \ and\ \bibinfo {author} {\bibfnamefont {T.~T.}\ \bibnamefont {Heikkil\"a}},\
  }\bibfield  {title} {\enquote {\bibinfo {title} {Flat-band superconductivity
  in strained dirac materials},}\ }\href {\doibase 10.1103/PhysRevB.93.214505}
  {\bibfield  {journal} {\bibinfo  {journal} {Phys. Rev. B}\ }\textbf {\bibinfo
  {volume} {93}},\ \bibinfo {pages} {214505} (\bibinfo {year}
  {2016})}\BibitemShut {NoStop}%
\bibitem [{\citenamefont {L\'opez-Sancho}\ and\ \citenamefont
  {Brey}(2016)}]{Brey2016}%
  \BibitemOpen
  \bibfield  {author} {\bibinfo {author} {\bibfnamefont {M.~Pilar}\
  \bibnamefont {L\'opez-Sancho}}\ and\ \bibinfo {author} {\bibfnamefont {Luis}\
  \bibnamefont {Brey}},\ }\bibfield  {title} {\enquote {\bibinfo {title}
  {Magnetic phases in periodically rippled graphene},}\ }\href {\doibase
  10.1103/PhysRevB.94.165430} {\bibfield  {journal} {\bibinfo  {journal} {Phys.
  Rev. B}\ }\textbf {\bibinfo {volume} {94}},\ \bibinfo {pages} {165430}
  (\bibinfo {year} {2016})}\BibitemShut {NoStop}%
\bibitem [{\citenamefont {Tang}\ \emph {et~al.}(2015)\citenamefont {Tang},
  \citenamefont {Laksono}, \citenamefont {Rodrigues}, \citenamefont {Sengupta},
  \citenamefont {Assaad},\ and\ \citenamefont {Adam}}]{Tang2015}%
  \BibitemOpen
  \bibfield  {author} {\bibinfo {author} {\bibfnamefont {Ho-Kin}\ \bibnamefont
  {Tang}}, \bibinfo {author} {\bibfnamefont {E.}~\bibnamefont {Laksono}},
  \bibinfo {author} {\bibfnamefont {J.~N.~B.}\ \bibnamefont {Rodrigues}},
  \bibinfo {author} {\bibfnamefont {P.}~\bibnamefont {Sengupta}}, \bibinfo
  {author} {\bibfnamefont {F.~F.}\ \bibnamefont {Assaad}}, \ and\ \bibinfo
  {author} {\bibfnamefont {S.}~\bibnamefont {Adam}},\ }\bibfield  {title}
  {\enquote {\bibinfo {title} {Interaction-driven metal-insulator transition in
  strained graphene},}\ }\href {\doibase 10.1103/PhysRevLett.115.186602}
  {\bibfield  {journal} {\bibinfo  {journal} {Phys. Rev. Lett.}\ }\textbf
  {\bibinfo {volume} {115}},\ \bibinfo {pages} {186602} (\bibinfo {year}
  {2015})}\BibitemShut {NoStop}%
\bibitem [{\citenamefont {Bae}\ \emph {et~al.}(2013)\citenamefont {Bae},
  \citenamefont {Lee}, \citenamefont {Sharma}, \citenamefont {Lee},
  \citenamefont {Kim},\ and\ \citenamefont {Ahn}}]{Bae2013}%
  \BibitemOpen
  \bibfield  {author} {\bibinfo {author} {\bibfnamefont {Sang-Hoon}\
  \bibnamefont {Bae}}, \bibinfo {author} {\bibfnamefont {Youngbin}\
  \bibnamefont {Lee}}, \bibinfo {author} {\bibfnamefont {Bhupendra~K.}\
  \bibnamefont {Sharma}}, \bibinfo {author} {\bibfnamefont {Hak-Joo}\
  \bibnamefont {Lee}}, \bibinfo {author} {\bibfnamefont {Jae-Hyun}\
  \bibnamefont {Kim}}, \ and\ \bibinfo {author} {\bibfnamefont {Jong-Hyun}\
  \bibnamefont {Ahn}},\ }\bibfield  {title} {\enquote {\bibinfo {title}
  {Graphene-based transparent strain sensor},}\ }\href {\doibase
  http://dx.doi.org/10.1016/j.carbon.2012.08.048} {\bibfield  {journal}
  {\bibinfo  {journal} {Carbon}\ }\textbf {\bibinfo {volume} {51}},\ \bibinfo
  {pages} {236 -- 242} (\bibinfo {year} {2013})}\BibitemShut {NoStop}%
\bibitem [{\citenamefont {Ni}\ \emph {et~al.}(2014)\citenamefont {Ni},
  \citenamefont {Yang}, \citenamefont {Ji}, \citenamefont {Baeck},
  \citenamefont {Toh}, \citenamefont {Ahn}, \citenamefont {Pereira},\ and\
  \citenamefont {\"Ozyilmaz}}]{Ni2014}%
  \BibitemOpen
  \bibfield  {author} {\bibinfo {author} {\bibfnamefont {Guang-Xin}\
  \bibnamefont {Ni}}, \bibinfo {author} {\bibfnamefont {Hong-Zhi}\ \bibnamefont
  {Yang}}, \bibinfo {author} {\bibfnamefont {Wei}\ \bibnamefont {Ji}}, \bibinfo
  {author} {\bibfnamefont {Seung-Jae}\ \bibnamefont {Baeck}}, \bibinfo {author}
  {\bibfnamefont {Chee-Tat}\ \bibnamefont {Toh}}, \bibinfo {author}
  {\bibfnamefont {Jong-Hyun}\ \bibnamefont {Ahn}}, \bibinfo {author}
  {\bibfnamefont {Vitor~M.}\ \bibnamefont {Pereira}}, \ and\ \bibinfo {author}
  {\bibfnamefont {Barbaros}\ \bibnamefont {\"Ozyilmaz}},\ }\bibfield  {title}
  {\enquote {\bibinfo {title} {Tuning optical conductivity of large-scale cvd
  graphene by strain engineering},}\ }\href {\doibase 10.1002/adma.201304156}
  {\bibfield  {journal} {\bibinfo  {journal} {Advanced Materials}\ }\textbf
  {\bibinfo {volume} {26}},\ \bibinfo {pages} {1081--1086} (\bibinfo {year}
  {2014})}\BibitemShut {NoStop}%
\bibitem [{\citenamefont {Chen}\ \emph {et~al.}(2016)\citenamefont {Chen},
  \citenamefont {Luo}, \citenamefont {Chen}, \citenamefont {Yan}, \citenamefont
  {Xu},\ and\ \citenamefont {Lu}}]{Chen2016}%
  \BibitemOpen
  \bibfield  {author} {\bibinfo {author} {\bibfnamefont {Jin-Hui}\ \bibnamefont
  {Chen}}, \bibinfo {author} {\bibfnamefont {Wei}\ \bibnamefont {Luo}},
  \bibinfo {author} {\bibfnamefont {Zhao-Xian}\ \bibnamefont {Chen}}, \bibinfo
  {author} {\bibfnamefont {Shao-Cheng}\ \bibnamefont {Yan}}, \bibinfo {author}
  {\bibfnamefont {Fei}\ \bibnamefont {Xu}}, \ and\ \bibinfo {author}
  {\bibfnamefont {Yan-Qing}\ \bibnamefont {Lu}},\ }\bibfield  {title} {\enquote
  {\bibinfo {title} {{Mechanical Modulation of a Hybrid Graphene--Microfiber
  Structure}},}\ }\href {\doibase 10.1002/adom.201500680} {\bibfield  {journal}
  {\bibinfo  {journal} {Advanced Optical Materials}\ }\textbf {\bibinfo
  {volume} {4}},\ \bibinfo {pages} {853--857} (\bibinfo {year}
  {2016})}\BibitemShut {NoStop}%
\bibitem [{\citenamefont {Rakheja}\ and\ \citenamefont
  {Sengupta}(2016)}]{Rakheja2016}%
  \BibitemOpen
  \bibfield  {author} {\bibinfo {author} {\bibfnamefont {Shaloo}\ \bibnamefont
  {Rakheja}}\ and\ \bibinfo {author} {\bibfnamefont {Parijat}\ \bibnamefont
  {Sengupta}},\ }\bibfield  {title} {\enquote {\bibinfo {title} {The tuning of
  light-matter coupling and dichroism in graphene for enhanced absorption:
  Implications for graphene-based optical absorption devices},}\ }\href
  {http://stacks.iop.org/0022-3727/49/i=11/a=115106} {\bibfield  {journal}
  {\bibinfo  {journal} {Journal of Physics D: Applied Physics}\ }\textbf
  {\bibinfo {volume} {49}},\ \bibinfo {pages} {115106} (\bibinfo {year}
  {2016})}\BibitemShut {NoStop}%
\bibitem [{\citenamefont {Nair}\ \emph {et~al.}(2008)\citenamefont {Nair},
  \citenamefont {Blake}, \citenamefont {Grigorenko}, \citenamefont {Novoselov},
  \citenamefont {Booth}, \citenamefont {Stauber}, \citenamefont {Peres},\ and\
  \citenamefont {Geim}}]{Nair2008}%
  \BibitemOpen
  \bibfield  {author} {\bibinfo {author} {\bibfnamefont {R.~R.}\ \bibnamefont
  {Nair}}, \bibinfo {author} {\bibfnamefont {P.}~\bibnamefont {Blake}},
  \bibinfo {author} {\bibfnamefont {A.~N.}\ \bibnamefont {Grigorenko}},
  \bibinfo {author} {\bibfnamefont {K.~S.}\ \bibnamefont {Novoselov}}, \bibinfo
  {author} {\bibfnamefont {T.~J.}\ \bibnamefont {Booth}}, \bibinfo {author}
  {\bibfnamefont {T.}~\bibnamefont {Stauber}}, \bibinfo {author} {\bibfnamefont
  {N.~M.~R.}\ \bibnamefont {Peres}}, \ and\ \bibinfo {author} {\bibfnamefont
  {A.~K.}\ \bibnamefont {Geim}},\ }\bibfield  {title} {\enquote {\bibinfo
  {title} {Fine structure constant defines visual transparency of graphene},}\
  }\href {\doibase 10.1126/science.1156965} {\bibfield  {journal} {\bibinfo
  {journal} {Science}\ }\textbf {\bibinfo {volume} {320}},\ \bibinfo {pages}
  {1308} (\bibinfo {year} {2008})}\BibitemShut {NoStop}%
\bibitem [{\citenamefont {Oliva-Leyva}\ and\ \citenamefont
  {Naumis}(2013)}]{Oliva13}%
  \BibitemOpen
  \bibfield  {author} {\bibinfo {author} {\bibfnamefont {M.}~\bibnamefont
  {Oliva-Leyva}}\ and\ \bibinfo {author} {\bibfnamefont {Gerardo~G.}\
  \bibnamefont {Naumis}},\ }\bibfield  {title} {\enquote {\bibinfo {title}
  {Understanding electron behavior in strained graphene as a reciprocal space
  distortion},}\ }\href {\doibase 10.1103/PhysRevB.88.085430} {\bibfield
  {journal} {\bibinfo  {journal} {Phys. Rev. B}\ }\textbf {\bibinfo {volume}
  {88}},\ \bibinfo {pages} {085430} (\bibinfo {year} {2013})}\BibitemShut
  {NoStop}%
\bibitem [{\citenamefont {Pellegrino}\ \emph {et~al.}(2010)\citenamefont
  {Pellegrino}, \citenamefont {Angilella},\ and\ \citenamefont
  {Pucci}}]{Pellegrino2010}%
  \BibitemOpen
  \bibfield  {author} {\bibinfo {author} {\bibfnamefont {F.~M.~D.}\
  \bibnamefont {Pellegrino}}, \bibinfo {author} {\bibfnamefont {G.~G.~N.}\
  \bibnamefont {Angilella}}, \ and\ \bibinfo {author} {\bibfnamefont
  {R.}~\bibnamefont {Pucci}},\ }\bibfield  {title} {\enquote {\bibinfo {title}
  {{Strain effect on the optical conductivity of graphene}},}\ }\href {\doibase
  10.1103/PhysRevB.81.035411} {\bibfield  {journal} {\bibinfo  {journal} {Phys.
  Rev. B}\ }\textbf {\bibinfo {volume} {81}},\ \bibinfo {pages} {035411}
  (\bibinfo {year} {2010})}\BibitemShut {NoStop}%
\bibitem [{\citenamefont {Pereira}\ \emph {et~al.}(2010)\citenamefont
  {Pereira}, \citenamefont {Ribeiro}, \citenamefont {Peres},\ and\
  \citenamefont {Neto}}]{Pereira2010}%
  \BibitemOpen
  \bibfield  {author} {\bibinfo {author} {\bibfnamefont {Vitor~M.}\
  \bibnamefont {Pereira}}, \bibinfo {author} {\bibfnamefont {R.~M.}\
  \bibnamefont {Ribeiro}}, \bibinfo {author} {\bibfnamefont {N.~M.~R.}\
  \bibnamefont {Peres}}, \ and\ \bibinfo {author} {\bibfnamefont
  {A.~H.~Castro}\ \bibnamefont {Neto}},\ }\bibfield  {title} {\enquote
  {\bibinfo {title} {{Optical properties of strained graphene}},}\ }\href
  {http://stacks.iop.org/0295-5075/92/i=6/a=67001} {\bibfield  {journal}
  {\bibinfo  {journal} {EPL (Europhysics Letters)}\ }\textbf {\bibinfo {volume}
  {92}},\ \bibinfo {pages} {67001} (\bibinfo {year} {2010})}\BibitemShut
  {NoStop}%
\bibitem [{\citenamefont {Oliva-Leyva}\ and\ \citenamefont
  {Naumis}(2014{\natexlab{a}})}]{Oliva2014}%
  \BibitemOpen
  \bibfield  {author} {\bibinfo {author} {\bibfnamefont {M.}~\bibnamefont
  {Oliva-Leyva}}\ and\ \bibinfo {author} {\bibfnamefont {Gerardo~G.}\
  \bibnamefont {Naumis}},\ }\bibfield  {title} {\enquote {\bibinfo {title}
  {{Anisotropic AC conductivity of strained graphene}},}\ }\href
  {http://stacks.iop.org/0953-8984/26/i=12/a=125302} {\bibfield  {journal}
  {\bibinfo  {journal} {Journal of Physics: Condensed Matter}\ }\textbf
  {\bibinfo {volume} {26}},\ \bibinfo {pages} {125302} (\bibinfo {year}
  {2014}{\natexlab{a}})}\BibitemShut {NoStop}%
\bibitem [{\citenamefont {Oliva-Leyva}\ and\ \citenamefont
  {Naumis}(2014{\natexlab{b}})}]{Oliva2014C}%
  \BibitemOpen
  \bibfield  {author} {\bibinfo {author} {\bibfnamefont {M.}~\bibnamefont
  {Oliva-Leyva}}\ and\ \bibinfo {author} {\bibfnamefont {Gerardo~G.}\
  \bibnamefont {Naumis}},\ }\bibfield  {title} {\enquote {\bibinfo {title}
  {{Corrigendum: Anisotropic AC conductivity of strained graphene (2014 J.
  Phys.: Condens. Matter 26 125302)}},}\ }\href
  {http://stacks.iop.org/0953-8984/26/i=27/a=279501} {\bibfield  {journal}
  {\bibinfo  {journal} {Journal of Physics: Condensed Matter}\ }\textbf
  {\bibinfo {volume} {26}},\ \bibinfo {pages} {279501} (\bibinfo {year}
  {2014}{\natexlab{b}})}\BibitemShut {NoStop}%
\bibitem [{\citenamefont {Oliva-Leyva}\ and\ \citenamefont
  {Naumis}(2015{\natexlab{a}})}]{Oliva2015}%
  \BibitemOpen
  \bibfield  {author} {\bibinfo {author} {\bibfnamefont {M}~\bibnamefont
  {Oliva-Leyva}}\ and\ \bibinfo {author} {\bibfnamefont {Gerardo~G}\
  \bibnamefont {Naumis}},\ }\bibfield  {title} {\enquote {\bibinfo {title}
  {Tunable dichroism and optical absorption of graphene by strain
  engineering},}\ }\href {http://stacks.iop.org/2053-1583/2/i=2/a=025001}
  {\bibfield  {journal} {\bibinfo  {journal} {2D Materials}\ }\textbf {\bibinfo
  {volume} {2}},\ \bibinfo {pages} {025001} (\bibinfo {year}
  {2015}{\natexlab{a}})}\BibitemShut {NoStop}%
\bibitem [{\citenamefont {P\'erez~Garza}\ \emph {et~al.}(2014)\citenamefont
  {P\'erez~Garza}, \citenamefont {Kievit}, \citenamefont {Schneider},\ and\
  \citenamefont {Staufer}}]{Perez2014}%
  \BibitemOpen
  \bibfield  {author} {\bibinfo {author} {\bibfnamefont {H.~Hugo}\ \bibnamefont
  {P\'erez~Garza}}, \bibinfo {author} {\bibfnamefont {Eric~W.}\ \bibnamefont
  {Kievit}}, \bibinfo {author} {\bibfnamefont {Gr\'egory~F.}\ \bibnamefont
  {Schneider}}, \ and\ \bibinfo {author} {\bibfnamefont {Urs}\ \bibnamefont
  {Staufer}},\ }\bibfield  {title} {\enquote {\bibinfo {title} {Controlled,
  reversible, and nondestructive generation of uniaxial extreme strains
  ($>10\%$) in graphene},}\ }\href {\doibase 10.1021/nl5016848} {\bibfield
  {journal} {\bibinfo  {journal} {Nano Letters}\ }\textbf {\bibinfo {volume}
  {14}},\ \bibinfo {pages} {4107--4113} (\bibinfo {year} {2014})}\BibitemShut
  {NoStop}%
\bibitem [{\citenamefont {Masir}\ \emph {et~al.}(2013)\citenamefont {Masir},
  \citenamefont {Moldovan},\ and\ \citenamefont {Peeters}}]{Ramezani2013}%
  \BibitemOpen
  \bibfield  {author} {\bibinfo {author} {\bibfnamefont {M.~Ramezani}\
  \bibnamefont {Masir}}, \bibinfo {author} {\bibfnamefont {D.}~\bibnamefont
  {Moldovan}}, \ and\ \bibinfo {author} {\bibfnamefont {F.M.}\ \bibnamefont
  {Peeters}},\ }\bibfield  {title} {\enquote {\bibinfo {title} {Pseudo magnetic
  field in strained graphene: Revisited},}\ }\href
  {http://www.sciencedirect.com/science/article/pii/S0038109813001555}
  {\bibfield  {journal} {\bibinfo  {journal} {Solid State Communications}\
  }\textbf {\bibinfo {volume} {175–176}},\ \bibinfo {pages} {76 -- 82}
  (\bibinfo {year} {2013})}\BibitemShut {NoStop}%
\bibitem [{\citenamefont {Li}(2014)}]{Li2014}%
  \BibitemOpen
  \bibfield  {author} {\bibinfo {author} {\bibfnamefont {Cui-Lian}\
  \bibnamefont {Li}},\ }\bibfield  {title} {\enquote {\bibinfo {title} {New
  position of dirac points in the strained graphene reciprocal lattice},}\
  }\href
  {http://scitation.aip.org/content/aip/journal/adva/4/8/10.1063/1.4893239}
  {\bibfield  {journal} {\bibinfo  {journal} {AIP Advances}\ }\textbf {\bibinfo
  {volume} {4}},\ \bibinfo {pages} {087119} (\bibinfo {year}
  {2014})}\BibitemShut {NoStop}%
\bibitem [{\citenamefont {Crosse}(2014)}]{Crosse2014}%
  \BibitemOpen
  \bibfield  {author} {\bibinfo {author} {\bibfnamefont {J.~A.}\ \bibnamefont
  {Crosse}},\ }\bibfield  {title} {\enquote {\bibinfo {title}
  {Large-displacement strain theory and its application to graphene},}\ }\href
  {\doibase 10.1103/PhysRevB.90.045201} {\bibfield  {journal} {\bibinfo
  {journal} {Phys. Rev. B}\ }\textbf {\bibinfo {volume} {90}},\ \bibinfo
  {pages} {045201} (\bibinfo {year} {2014})}\BibitemShut {NoStop}%
\bibitem [{\citenamefont {Ray}\ \emph {et~al.}(2016)\citenamefont {Ray},
  \citenamefont {Rost}, \citenamefont {Weckbecker}, \citenamefont {Vogl},
  \citenamefont {Sharma}, \citenamefont {Gupta}, \citenamefont {Pankratov},\
  and\ \citenamefont {Shallcross}}]{Shallcross2016}%
  \BibitemOpen
  \bibfield  {author} {\bibinfo {author} {\bibfnamefont {N.}~\bibnamefont
  {Ray}}, \bibinfo {author} {\bibfnamefont {F.}~\bibnamefont {Rost}}, \bibinfo
  {author} {\bibfnamefont {D.}~\bibnamefont {Weckbecker}}, \bibinfo {author}
  {\bibfnamefont {M.}~\bibnamefont {Vogl}}, \bibinfo {author} {\bibfnamefont
  {S.}~\bibnamefont {Sharma}}, \bibinfo {author} {\bibfnamefont
  {R.}~\bibnamefont {Gupta}}, \bibinfo {author} {\bibfnamefont
  {O.}~\bibnamefont {Pankratov}}, \ and\ \bibinfo {author} {\bibfnamefont
  {S.}~\bibnamefont {Shallcross}},\ }\bibfield  {title} {\enquote {\bibinfo
  {title} {{Going beyond k.p theory: a general method for obtaining effective
  Hamiltonians in both high and low symmetry situations}},}\ }\href
  {https://arxiv.org/abs/1607.00920} {\bibfield  {journal} {\bibinfo  {journal}
  {arXiv:1607.00920}\ } (\bibinfo {year} {2016})}\BibitemShut {NoStop}%
\bibitem [{\citenamefont {Qi}\ \emph {et~al.}(2013)\citenamefont {Qi},
  \citenamefont {Bahamon}, \citenamefont {Pereira}, \citenamefont {Park},
  \citenamefont {Campbell},\ and\ \citenamefont {Neto}}]{Zenan13}%
  \BibitemOpen
  \bibfield  {author} {\bibinfo {author} {\bibfnamefont {Zenan}\ \bibnamefont
  {Qi}}, \bibinfo {author} {\bibfnamefont {D.~A.}\ \bibnamefont {Bahamon}},
  \bibinfo {author} {\bibfnamefont {Vitor~M.}\ \bibnamefont {Pereira}},
  \bibinfo {author} {\bibfnamefont {Harold~S.}\ \bibnamefont {Park}}, \bibinfo
  {author} {\bibfnamefont {D.~K.}\ \bibnamefont {Campbell}}, \ and\ \bibinfo
  {author} {\bibfnamefont {A.~H.~Castro}\ \bibnamefont {Neto}},\ }\bibfield
  {title} {\enquote {\bibinfo {title} {Resonant tunneling in graphene
  pseudomagnetic quantum dots},}\ }\href {\doibase 10.1021/nl400872q}
  {\bibfield  {journal} {\bibinfo  {journal} {Nano Letters}\ }\textbf {\bibinfo
  {volume} {13}},\ \bibinfo {pages} {2692--2697} (\bibinfo {year}
  {2013})}\BibitemShut {NoStop}%
\bibitem [{\citenamefont {Sloan}\ \emph {et~al.}(2013)\citenamefont {Sloan},
  \citenamefont {Sanjuan}, \citenamefont {Wang}, \citenamefont {Horvath},\ and\
  \citenamefont {Barraza-Lopez}}]{Salvador13}%
  \BibitemOpen
  \bibfield  {author} {\bibinfo {author} {\bibfnamefont {James~V.}\
  \bibnamefont {Sloan}}, \bibinfo {author} {\bibfnamefont {Alejandro
  A.~Pacheco}\ \bibnamefont {Sanjuan}}, \bibinfo {author} {\bibfnamefont
  {Zhengfei}\ \bibnamefont {Wang}}, \bibinfo {author} {\bibfnamefont {Cedric}\
  \bibnamefont {Horvath}}, \ and\ \bibinfo {author} {\bibfnamefont {Salvador}\
  \bibnamefont {Barraza-Lopez}},\ }\bibfield  {title} {\enquote {\bibinfo
  {title} {Strain gauge fields for rippled graphene membranes under central
  mechanical load: An approach beyond first-order continuum elasticity},}\
  }\href {\doibase 10.1103/PhysRevB.87.155436} {\bibfield  {journal} {\bibinfo
  {journal} {Phys. Rev. B}\ }\textbf {\bibinfo {volume} {87}},\ \bibinfo
  {pages} {155436} (\bibinfo {year} {2013})}\BibitemShut {NoStop}%
\bibitem [{\citenamefont {Papaconstantopoulos}\ \emph
  {et~al.}(1998)\citenamefont {Papaconstantopoulos}, \citenamefont {Mehl},
  \citenamefont {Erwin},\ and\ \citenamefont {Pederson}}]{Papaconstantopoulos}%
  \BibitemOpen
  \bibfield  {author} {\bibinfo {author} {\bibfnamefont {D.~A.}\ \bibnamefont
  {Papaconstantopoulos}}, \bibinfo {author} {\bibfnamefont {M.~J.}\
  \bibnamefont {Mehl}}, \bibinfo {author} {\bibfnamefont {S.~C.}\ \bibnamefont
  {Erwin}}, \ and\ \bibinfo {author} {\bibfnamefont {M.~R.}\ \bibnamefont
  {Pederson}},\ }\bibfield  {title} {\enquote {\bibinfo {title} {Tight-binding
  hamiltonians for carbon and silicon},}\ }in\ \href@noop {} {\emph {\bibinfo
  {booktitle} {Tight-Binding Approach to Computational Materials Science}}},\
  \bibinfo {editor} {edited by\ \bibinfo {editor} {\bibfnamefont
  {P.}~\bibnamefont {Turchi}}, \bibinfo {editor} {\bibfnamefont
  {A.}~\bibnamefont {Gonis}}, \ and\ \bibinfo {editor} {\bibfnamefont
  {L.}~\bibnamefont {Colombo}}}\ (\bibinfo  {publisher} {Materials Research
  Society},\ \bibinfo {address} {Pittsburgh},\ \bibinfo {year} {1998})\ p.\
  \bibinfo {pages} {221}\BibitemShut {NoStop}%
\bibitem [{\citenamefont {Ribeiro}\ \emph {et~al.}(2009)\citenamefont
  {Ribeiro}, \citenamefont {Pereira}, \citenamefont {Peres}, \citenamefont
  {Briddon},\ and\ \citenamefont {Castro~Neto}}]{Ribeiro09}%
  \BibitemOpen
  \bibfield  {author} {\bibinfo {author} {\bibfnamefont {R.~M.}\ \bibnamefont
  {Ribeiro}}, \bibinfo {author} {\bibfnamefont {Vitor~M.}\ \bibnamefont
  {Pereira}}, \bibinfo {author} {\bibfnamefont {N.~M.~R.}\ \bibnamefont
  {Peres}}, \bibinfo {author} {\bibfnamefont {P.~R.}\ \bibnamefont {Briddon}},
  \ and\ \bibinfo {author} {\bibfnamefont {A.~H.}\ \bibnamefont
  {Castro~Neto}},\ }\bibfield  {title} {\enquote {\bibinfo {title} {Strained
  graphene: tight-binding and density functional calculations},}\ }\href
  {http://stacks.iop.org/1367-2630/11/i=11/a=115002} {\bibfield  {journal}
  {\bibinfo  {journal} {New J. Phys.}\ }\textbf {\bibinfo {volume} {11}},\
  \bibinfo {pages} {115002} (\bibinfo {year} {2009})}\BibitemShut {NoStop}%
\bibitem [{\citenamefont {Choi}\ \emph
  {et~al.}(2010{\natexlab{b}})\citenamefont {Choi}, \citenamefont {Jhi},\ and\
  \citenamefont {Son}}]{Choi10}%
  \BibitemOpen
  \bibfield  {author} {\bibinfo {author} {\bibfnamefont {Seon-Myeong}\
  \bibnamefont {Choi}}, \bibinfo {author} {\bibfnamefont {Seung-Hoon}\
  \bibnamefont {Jhi}}, \ and\ \bibinfo {author} {\bibfnamefont {Young-Woo}\
  \bibnamefont {Son}},\ }\bibfield  {title} {\enquote {\bibinfo {title}
  {Effects of strain on electronic properties of graphene},}\ }\href {\doibase
  10.1103/PhysRevB.81.081407} {\bibfield  {journal} {\bibinfo  {journal} {Phys.
  Rev. B}\ }\textbf {\bibinfo {volume} {81}},\ \bibinfo {pages} {081407}
  (\bibinfo {year} {2010}{\natexlab{b}})}\BibitemShut {NoStop}%
\bibitem [{\citenamefont {Ni}\ \emph {et~al.}(2009)\citenamefont {Ni},
  \citenamefont {Yu}, \citenamefont {Lu}, \citenamefont {Wang}, \citenamefont
  {Feng},\ and\ \citenamefont {Shen}}]{Ni09}%
  \BibitemOpen
  \bibfield  {author} {\bibinfo {author} {\bibfnamefont {Zhen~Hua}\
  \bibnamefont {Ni}}, \bibinfo {author} {\bibfnamefont {Ting}\ \bibnamefont
  {Yu}}, \bibinfo {author} {\bibfnamefont {Yun~Hao}\ \bibnamefont {Lu}},
  \bibinfo {author} {\bibfnamefont {Ying~Ying}\ \bibnamefont {Wang}}, \bibinfo
  {author} {\bibfnamefont {Yuan~Ping}\ \bibnamefont {Feng}}, \ and\ \bibinfo
  {author} {\bibfnamefont {Ze~Xiang}\ \bibnamefont {Shen}},\ }\bibfield
  {title} {\enquote {\bibinfo {title} {Uniaxial strain on graphene: Raman
  spectroscopy study and band-gap opening},}\ }\href {\doibase
  10.1021/nn8008323} {\bibfield  {journal} {\bibinfo  {journal} {ACS Nano}\
  }\textbf {\bibinfo {volume} {3}},\ \bibinfo {pages} {483--483} (\bibinfo
  {year} {2009})}\BibitemShut {NoStop}%
\bibitem [{\citenamefont {Li}\ \emph {et~al.}(2010)\citenamefont {Li},
  \citenamefont {Jiang}, \citenamefont {Liu},\ and\ \citenamefont
  {Liu}}]{YangLi10}%
  \BibitemOpen
  \bibfield  {author} {\bibinfo {author} {\bibfnamefont {Yang}\ \bibnamefont
  {Li}}, \bibinfo {author} {\bibfnamefont {Xiaowei}\ \bibnamefont {Jiang}},
  \bibinfo {author} {\bibfnamefont {Zhongfan}\ \bibnamefont {Liu}}, \ and\
  \bibinfo {author} {\bibfnamefont {Zhirong}\ \bibnamefont {Liu}},\ }\bibfield
  {title} {\enquote {\bibinfo {title} {Strain effects in graphene and graphene
  nanoribbons: The underlying mechanism},}\ }\href {\doibase
  10.1007/s12274-010-0015-7} {\bibfield  {journal} {\bibinfo  {journal} {Nano
  Research}\ }\textbf {\bibinfo {volume} {3}},\ \bibinfo {pages} {545--556}
  (\bibinfo {year} {2010})}\BibitemShut {NoStop}%
\bibitem [{\citenamefont {Oliva-Leyva}\ and\ \citenamefont
  {Naumis}(2015{\natexlab{b}})}]{Oliva15b}%
  \BibitemOpen
  \bibfield  {author} {\bibinfo {author} {\bibfnamefont {M.}~\bibnamefont
  {Oliva-Leyva}}\ and\ \bibinfo {author} {\bibfnamefont {Gerardo~G.}\
  \bibnamefont {Naumis}},\ }\bibfield  {title} {\enquote {\bibinfo {title}
  {Generalizing the fermi velocity of strained graphene from uniform to
  nonuniform strain},}\ }\href {\doibase
  http://dx.doi.org/10.1016/j.physleta.2015.05.039} {\bibfield  {journal}
  {\bibinfo  {journal} {Physics Letters A}\ }\textbf {\bibinfo {volume}
  {379}},\ \bibinfo {pages} {2645 -- 2651} (\bibinfo {year}
  {2015}{\natexlab{b}})}\BibitemShut {NoStop}%
\bibitem [{\citenamefont {Volovik}\ and\ \citenamefont
  {Zubkov}(2014)}]{Volovik14a}%
  \BibitemOpen
  \bibfield  {author} {\bibinfo {author} {\bibfnamefont {G.E.}\ \bibnamefont
  {Volovik}}\ and\ \bibinfo {author} {\bibfnamefont {M.A.}\ \bibnamefont
  {Zubkov}},\ }\bibfield  {title} {\enquote {\bibinfo {title} {Emergent horava
  gravity in graphene},}\ }\href {\doibase
  http://dx.doi.org/10.1016/j.aop.2013.11.003} {\bibfield  {journal} {\bibinfo
  {journal} {Annals of Physics}\ }\textbf {\bibinfo {volume} {340}},\ \bibinfo
  {pages} {352 -- 368} (\bibinfo {year} {2014})}\BibitemShut {NoStop}%
\bibitem [{\citenamefont {Yang}(2011)}]{Yang2011}%
  \BibitemOpen
  \bibfield  {author} {\bibinfo {author} {\bibfnamefont {Hua~Tong}\
  \bibnamefont {Yang}},\ }\bibfield  {title} {\enquote {\bibinfo {title}
  {{Strain induced shift of Dirac points and the pseudo-magnetic field in
  graphene}},}\ }\href {http://stacks.iop.org/0953-8984/23/i=50/a=505502}
  {\bibfield  {journal} {\bibinfo  {journal} {Journal of Physics: Condensed
  Matter}\ }\textbf {\bibinfo {volume} {23}},\ \bibinfo {pages} {505502}
  (\bibinfo {year} {2011})}\BibitemShut {NoStop}%
\bibitem [{\citenamefont {Bahat-Treidel}\ \emph {et~al.}(2010)\citenamefont
  {Bahat-Treidel}, \citenamefont {Peleg}, \citenamefont {Grobman},
  \citenamefont {Shapira}, \citenamefont {Segev},\ and\ \citenamefont
  {Pereg-Barnea}}]{Bahat10}%
  \BibitemOpen
  \bibfield  {author} {\bibinfo {author} {\bibfnamefont {Omri}\ \bibnamefont
  {Bahat-Treidel}}, \bibinfo {author} {\bibfnamefont {Or}~\bibnamefont
  {Peleg}}, \bibinfo {author} {\bibfnamefont {Mark}\ \bibnamefont {Grobman}},
  \bibinfo {author} {\bibfnamefont {Nadav}\ \bibnamefont {Shapira}}, \bibinfo
  {author} {\bibfnamefont {Mordechai}\ \bibnamefont {Segev}}, \ and\ \bibinfo
  {author} {\bibfnamefont {T.}~\bibnamefont {Pereg-Barnea}},\ }\bibfield
  {title} {\enquote {\bibinfo {title} {Klein tunneling in deformed honeycomb
  lattices},}\ }\href {\doibase 10.1103/PhysRevLett.104.063901} {\bibfield
  {journal} {\bibinfo  {journal} {Phys. Rev. Lett.}\ }\textbf {\bibinfo
  {volume} {104}},\ \bibinfo {pages} {063901} (\bibinfo {year}
  {2010})}\BibitemShut {NoStop}%
\bibitem [{\citenamefont {Volovik}\ and\ \citenamefont
  {Zubkov}(2015)}]{Volovik15}%
  \BibitemOpen
  \bibfield  {author} {\bibinfo {author} {\bibfnamefont {G.E.}\ \bibnamefont
  {Volovik}}\ and\ \bibinfo {author} {\bibfnamefont {M.A.}\ \bibnamefont
  {Zubkov}},\ }\bibfield  {title} {\enquote {\bibinfo {title} {Emergent
  geometry experienced by fermions in graphene in the presence of
  dislocations},}\ }\href {\doibase
  http://dx.doi.org/10.1016/j.aop.2015.03.005} {\bibfield  {journal} {\bibinfo
  {journal} {Annals of Physics}\ }\textbf {\bibinfo {volume} {356}},\ \bibinfo
  {pages} {255 -- 268} (\bibinfo {year} {2015})}\BibitemShut {NoStop}%
\bibitem [{\citenamefont {Barber}(2003)}]{Barber}%
  \BibitemOpen
  \bibfield  {author} {\bibinfo {author} {\bibfnamefont {J.R.}\ \bibnamefont
  {Barber}},\ }\href@noop {} {\emph {\bibinfo {title} {Elasticity}}},\ \bibinfo
  {edition} {2nd}\ ed.,\ Solid Mechanics and Its Applications\ (\bibinfo
  {publisher} {Springer},\ \bibinfo {year} {2003})\ \bibinfo {note} {p.
  8}\BibitemShut {NoStop}%
\bibitem [{\citenamefont {Oliva-Leyva}\ and\ \citenamefont
  {Naumis}(2016)}]{Oliva2016}%
  \BibitemOpen
  \bibfield  {author} {\bibinfo {author} {\bibfnamefont {M.}~\bibnamefont
  {Oliva-Leyva}}\ and\ \bibinfo {author} {\bibfnamefont {Gerardo~G.}\
  \bibnamefont {Naumis}},\ }\bibfield  {title} {\enquote {\bibinfo {title}
  {Effective dirac hamiltonian for anisotropic honeycomb lattices: Optical
  properties},}\ }\href {\doibase 10.1103/PhysRevB.93.035439} {\bibfield
  {journal} {\bibinfo  {journal} {Phys. Rev. B}\ }\textbf {\bibinfo {volume}
  {93}},\ \bibinfo {pages} {035439} (\bibinfo {year} {2016})}\BibitemShut
  {NoStop}%
\bibitem [{\citenamefont {Ziegler}(2007)}]{Ziegler07}%
  \BibitemOpen
  \bibfield  {author} {\bibinfo {author} {\bibfnamefont {K.}~\bibnamefont
  {Ziegler}},\ }\bibfield  {title} {\enquote {\bibinfo {title} {Minimal
  conductivity of graphene: Nonuniversal values from the kubo formula},}\
  }\href {\doibase 10.1103/PhysRevB.75.233407} {\bibfield  {journal} {\bibinfo
  {journal} {Phys. Rev. B}\ }\textbf {\bibinfo {volume} {75}},\ \bibinfo
  {pages} {233407} (\bibinfo {year} {2007})}\BibitemShut {NoStop}%
\bibitem [{\citenamefont {Gusynin}\ \emph {et~al.}(2007)\citenamefont
  {Gusynin}, \citenamefont {Sharapov},\ and\ \citenamefont
  {Carbotte}}]{Gusynin07}%
  \BibitemOpen
  \bibfield  {author} {\bibinfo {author} {\bibfnamefont {V.~P.}\ \bibnamefont
  {Gusynin}}, \bibinfo {author} {\bibfnamefont {S.~G.}\ \bibnamefont
  {Sharapov}}, \ and\ \bibinfo {author} {\bibfnamefont {J.~P.}\ \bibnamefont
  {Carbotte}},\ }\bibfield  {title} {\enquote {\bibinfo {title} {Ac
  conductivity of graphene: From tight-binding model to 2 + 1-dimensional
  quantum electrodynamics},}\ }\href {\doibase 10.1142/S0217979207038022}
  {\bibfield  {journal} {\bibinfo  {journal} {International Journal of Modern
  Physics B}\ }\textbf {\bibinfo {volume} {21}},\ \bibinfo {pages} {4611--4658}
  (\bibinfo {year} {2007})}\BibitemShut {NoStop}%
\bibitem [{\citenamefont {Stauber}\ \emph {et~al.}(2015)\citenamefont
  {Stauber}, \citenamefont {Noriega-P\'erez},\ and\ \citenamefont
  {Schliemann}}]{Stauber2015}%
  \BibitemOpen
  \bibfield  {author} {\bibinfo {author} {\bibfnamefont {T.}~\bibnamefont
  {Stauber}}, \bibinfo {author} {\bibfnamefont {D.}~\bibnamefont
  {Noriega-P\'erez}}, \ and\ \bibinfo {author} {\bibfnamefont {J.}~\bibnamefont
  {Schliemann}},\ }\bibfield  {title} {\enquote {\bibinfo {title} {Universal
  absorption of two-dimensional systems},}\ }\href {\doibase
  10.1103/PhysRevB.91.115407} {\bibfield  {journal} {\bibinfo  {journal} {Phys.
  Rev. B}\ }\textbf {\bibinfo {volume} {91}},\ \bibinfo {pages} {115407}
  (\bibinfo {year} {2015})}\BibitemShut {NoStop}%
\bibitem [{\citenamefont {Stauber}\ \emph {et~al.}(2008)\citenamefont
  {Stauber}, \citenamefont {Peres},\ and\ \citenamefont {Geim}}]{Stauber2008}%
  \BibitemOpen
  \bibfield  {author} {\bibinfo {author} {\bibfnamefont {T.}~\bibnamefont
  {Stauber}}, \bibinfo {author} {\bibfnamefont {N.~M.~R.}\ \bibnamefont
  {Peres}}, \ and\ \bibinfo {author} {\bibfnamefont {A.~K.}\ \bibnamefont
  {Geim}},\ }\bibfield  {title} {\enquote {\bibinfo {title} {Optical
  conductivity of graphene in the visible region of the spectrum},}\ }\href
  {\doibase 10.1103/PhysRevB.78.085432} {\bibfield  {journal} {\bibinfo
  {journal} {Phys. Rev. B}\ }\textbf {\bibinfo {volume} {78}},\ \bibinfo
  {pages} {085432} (\bibinfo {year} {2008})}\BibitemShut {NoStop}%
\bibitem [{\citenamefont {Castro~Neto}\ \emph {et~al.}(2009)\citenamefont
  {Castro~Neto}, \citenamefont {Guinea}, \citenamefont {Peres}, \citenamefont
  {Novoselov},\ and\ \citenamefont {Geim}}]{Neto09}%
  \BibitemOpen
  \bibfield  {author} {\bibinfo {author} {\bibfnamefont {A.~H.}\ \bibnamefont
  {Castro~Neto}}, \bibinfo {author} {\bibfnamefont {F.}~\bibnamefont {Guinea}},
  \bibinfo {author} {\bibfnamefont {N.~M.~R.}\ \bibnamefont {Peres}}, \bibinfo
  {author} {\bibfnamefont {K.~S.}\ \bibnamefont {Novoselov}}, \ and\ \bibinfo
  {author} {\bibfnamefont {A.~K.}\ \bibnamefont {Geim}},\ }\bibfield  {title}
  {\enquote {\bibinfo {title} {The electronic properties of graphene},}\ }\href
  {\doibase 10.1103/RevModPhys.81.109} {\bibfield  {journal} {\bibinfo
  {journal} {Rev. Mod. Phys.}\ }\textbf {\bibinfo {volume} {81}},\ \bibinfo
  {pages} {109--162} (\bibinfo {year} {2009})}\BibitemShut {NoStop}%
\bibitem [{\citenamefont {Ziegler}(2006)}]{Ziegler06}%
  \BibitemOpen
  \bibfield  {author} {\bibinfo {author} {\bibfnamefont {K.}~\bibnamefont
  {Ziegler}},\ }\bibfield  {title} {\enquote {\bibinfo {title} {Robust
  transport properties in graphene},}\ }\href {\doibase
  10.1103/PhysRevLett.97.266802} {\bibfield  {journal} {\bibinfo  {journal}
  {Phys. Rev. Lett.}\ }\textbf {\bibinfo {volume} {97}},\ \bibinfo {pages}
  {266802} (\bibinfo {year} {2006})}\BibitemShut {NoStop}%
\end{thebibliography}%

\end{document}